\documentclass[a4paper,fleqn,usenatbib]{mnras}

\usepackage{txfonts}
\usepackage[authoryear]{natbib}
\usepackage{lscape}                                
\usepackage{color}
\usepackage{graphicx}
\usepackage{longtable}

\title[Neutron-capture elements in NGC\,5315]{Neutron-capture element abundances in the planetary nebula NGC\,5315 from deep optical and near-infrared spectrophotometry\thanks{This paper includes observations collected at the European Southern Observatory, Chile, proposal number ESO 092.D-0189(A).}\thanks{This paper includes data gathered with the 6.5 meter Magellan Telescopes located at Las Campanas Observatory, Chile.}}

\author[S. Madonna et al.]{
S. Madonna,$^{1,2}$\thanks{E-mail: smadonna@iac.es}
J. Garc\'{i}a-Rojas,$^{1,2}$
N. C. Sterling,$^{3}$
G. Delgado-Inglada,$^{4}$ 
\newauthor
A. Mesa-Delgado,$^{5}$
V. Luridiana,$^{1,2}$
I. U. Roederer$^{6,7}$
and A. L. Mashburn$^{3}$
\\
$^{1}$Instituto de Astrof{\'i}sica de Canarias, E-38205 La Laguna, Tenerife, Spain\\
$^{2}$Universidad de La Laguna, Dpto. Astrof{\'i}sica, E-38206 La Laguna, Tenerife, Spain\\
$^{3}$Department of Physics, University of West Georgia, 1601 Maple Street, Carrollton, GA 30118, USA\\
$^{4}$Instituto de Astronom\'{\i}a, Universidad Nacional Aut\'onoma de M\'exico, Apdo. Postal 70264, Ciudad de M\'exico, 04510, Mexico\\
$^{5}$Instituto de Astrof\'{\i}sica, Facultad de F\'{\i}sica, Pontificia Universidad Cat\'olica de Chile, Av. Vicu\~na Mackenna 4860,782-0436 Macul, Santiago, Chile\\
$^{6}$Department of Astronomy, University of Michigan, 1085 South University Avenue, Ann Arbor, MI 48109, USA\\
$^{7}$Joint Institute for Nuclear Astrophysics and Center for the Evolution of the Elements (JINA-CEE), USA\\
}

\date{Accepted XXX. Received YYY; in original form ZZZ}

\pubyear{2016}

\newcommand{\elecd}{$n_{\rm e}$}
\newcommand{\te}{$T_{\rm e}$}
\newcommand{\hb}{H$\beta$}

\newcommand{\fci}{[C~{\sc i}]}

\newcommand{\foi}{[O~{\sc i}]}
\newcommand{\foii}{[O~{\sc ii}]}
\newcommand{\foiii}{[O~{\sc iii}]}

\newcommand{\fsii}{[S~{\sc ii}]}
\newcommand{\fsiii}{[S~{\sc iii}]}
\newcommand{\fnitroi}{[N~{\sc i}]}
\newcommand{\fnii}{[N~{\sc ii}]}
\newcommand{\sfmgi}{Mg~{\sc i}]}

\newcommand{\mgii}{Mg~{\sc ii}}
\newcommand{\fariii}{[Ar~{\sc iii}]}
\newcommand{\fariv}{[Ar~{\sc iv}]}

\newcommand{\fclii}{[Cl~{\sc ii}]}
\newcommand{\fcliii}{[Cl~{\sc iii}]}
\newcommand{\fcliv}{[Cl~{\sc iv}]}
\newcommand{\fcrii}{[Cr~{\sc ii}]}
\newcommand{\fcriii}{[Cr~{\sc iii}]}

\newcommand{\fneiii}{[Ne~{\sc iii}]}
\newcommand{\fneiv}{[Ne~{\sc iv}]}
\newcommand{\fnev}{[Ne~{\sc v}]}
\newcommand{\fkriii}{[Kr~{\sc iii}]}
\newcommand{\fkriv}{[Kr~{\sc iv}]}

\newcommand{\fxeiii}{[Xe~{\sc iii}]}
\newcommand{\fxeiv}{[Xe~{\sc iv}]}

\newcommand{\frbiv}{[Rb~{\sc iv}]}

\newcommand{\fniqii}{[Ni~{\sc ii}]}

\newcommand{\fmniii}{[Mn~{\sc iii}]}

\newcommand{\ffeii}{[Fe~{\sc ii}]}
\newcommand{\ffeiii}{[Fe~{\sc iii}]}
\newcommand{\ffeiv}{[Fe~{\sc iv}]}

\newcommand{\fpii}{[P~{\sc ii}]}

\newcommand{\fseiii}{[Se~{\sc iii}]}
\newcommand{\fseiv}{[Se~{\sc iv}]}
\newcommand{\fbriii}{[Br~{\sc iii}]}
\newcommand{\oiii}{O~{\sc iii}}

\newcommand{\nitroi}{N~{\sc i}}
\newcommand{\nii}{N~{\sc ii}}
\newcommand{\niii}{N~{\sc iii}}
\newcommand{\niv}{N~{\sc iv}}

\newcommand{\silii}{Si~{\sc ii}}
\newcommand{\siliii}{Si~{\sc iii}}

\newcommand{\oi}{O~{\sc i}}
\newcommand{\oii}{O~{\sc ii}}
\newcommand{\ci}{C~{\sc i}}
\newcommand{\cii}{C~{\sc ii}}
\newcommand{\ciii}{C~{\sc iii}}
\newcommand{\civ}{C~{\sc iv}}

\newcommand{\nei}{Ne~{\sc i}}
\newcommand{\neii}{Ne~{\sc ii}}

\newcommand{\sii}{S~{\sc ii}}
\newcommand{\siii}{S~{\sc iii}}

\newcommand{\niqii}{Ni~{\sc ii}}

\newcommand{\arii}{Ar~{\sc ii}}

\newcommand{\hi}{H\,{\sc i}}
\newcommand{\hii}{H~{\sc ii}}

\newcommand{\hei}{He~{\sc i}}
\newcommand{\heii}{He~{\sc ii}}

\begin{document}
\label{firstpage}
\pagerange{\pageref{firstpage}--\pageref{lastpage}}
\maketitle

\begin{abstract}
We analyze the chemical composition of the planetary nebula (PN) NGC\,5315, through high-resolution (R$\sim$40000) optical spectroscopy with UVES at the Very Large Telescope, and medium-resolution (R$\sim$4800) near-infrared spectroscopy with FIRE at Magellan Baade Telescope, covering a wide spectral range from 0.31 to 2.50 $\mu$m. The main aim of this work is to investigate neutron (\emph{n})-capture element abundances to study the operation of the slow \emph{n}-capture (``\emph{s}-process'') in the AGB progenitor of NGC\,5315. We detect more than 700 emission lines, including ions of the \emph{n}-capture elements Se, Kr, Xe, and possibly Br. We compute physical conditions from a large number of diagnostic line ratios, and derive ionic abundances for species with available atomic data. The total abundances are computed using recent ionization correction factors (ICFs) or by summing ionic abundances. Total abundances of common elements are in good agreement with previous work on this object.
Based on our abundance analysis of NGC\,5315, including the lack of \emph{s}-process enrichment, we  speculate that the most probable scenario is that the progenitor star is in a binary system as hinted at by radial velocity studies, and interactions with its companion truncated the AGB before \emph{s}-process enrichment could occur. However there are other two possible scenarios for its evolution, that cannot be ruled out: i) the progenitor is a low-mass single star that did not undergo third dredge-up; ii) the progenitor star of NGC\,5315 had an initial mass of 4--6 M$_{\odot}$, and any \emph{s}-process enhancements were heavily diluted by the massive envelope during the AGB phase.  
\end{abstract}




\section{Introduction}\label{sec:intro}

Trans-iron (\emph{n}-capture) elements are produced in asymptotic giant branch (AGB) stars (1--8 M$_{\odot}$) in the intershell region between the H- and He-burning shells, in the so-called \emph{s}-process (slow neutron-capture process). 
In these layers neutrons are released by $\alpha$-captures onto $^{13}$C (or $^{22}$Ne in AGB stars with mass $>3-4$ M$_{\odot}$). During the thermally-pulsing phase, convective dredge-up conveys to the stellar surface C and \emph{s}-process-enriched material, which is then expelled to the interstellar medium in the planetary nebula (hereafter PN) phase and eventually incorporated in a new generation of stars. Thus, the abundances of \emph{n}-capture elements and the \emph{s}-process element-by-element enrichment pattern reveal critical information on physical conditions in stellar interiors and the nucleosynthetic histories of stellar populations \citep[][]{cristalloetal11, cristalloetal15, karakaslattanzio14, trippellaetal14, trippellaetal16, venturaetal15}.

While \emph{s}-process nucleosynthesis has historically been studied through stellar spectra, nebular spectroscopy presents unique advantages. PNe allow for the first observational analysis of the lightest {\emph n}-capture elements (Ge, Se, and Br) and the noble gases Kr and Xe in one of their sites of origin.  A crucial difference from AGB stars is that PNe represent the final envelope abundances of their progenitor stars, after the cessation of nucleosynthesis and dredge-up that occurred during the AGB phase. PN abundances are therefore highly valuable for stellar yield determinations.  Moreover, nebular abundance determinations provide key constraints to poorly-understood processes in models of stellar evolution and nucleosynthesis, such as the efficiency of third dredge-up (TDU) at low envelope masses, the number of thermal pulses, and treatments of mass loss and convective overshoot \citep[][and references therein]{cristalloetal15, karakasetal09, karakasetal12, karakaslattanzio14, trippellaetal16, venturaetal15}.  The number of {\emph n}-capture elements that can be detected in individual PNe allows for meaningful comparisons with various sets of evolutionary models \citep[e.~g.,][hereafter S16]{sterlingetal16}.  In particular, the optical region is home to numerous {\emph n}-capture element transitions, including lines from multiple ions of Se, Br, Kr, Rb, and Xe that have been detected in various PNe \citep[e.~g. ][]{sharpeeetal07, garciarojasetal12, garciarojasetal15}.

Since the pioneering work by \citet{pequignotbaluteau94}, only a few detailed studies of \emph{n}-capture element abundances based on deep, high-resolution optical spectra  have been conducted in PNe. \citet{sharpeeetal07} identified lines of Br, Kr, Rb, Xe, Ba, and possibly Te and I in 4 PNe, with 4- and 6-m class telescopes. But at their resolution of $\sim$22,000, many features were not unambiguously detected. \citet{garciarojasetal15} made a detailed {\emph n}-capture element abundance analysis in NGC\,3918, a PN with a high ionization degree and with a C/O ratio close to 1, detecting several ions of Kr, Xe, Rb and Se. This allowed them to test the new ionization correction factors (ICFs) for \emph{n}-capture elements provided by \citet{sterlingetal15} and to compute total abundances with unprecedented accuracy. In contrast, near-infrared (hereafter NIR) lines of \emph{n}-capture elements have been studied in a large number of PNe. For example, \citet{sterlingdinerstein08}'s survey of {\fkriii} and {\fseiv} in 120 PNe resulted in the first overview of \emph{s}-process enrichments in PNe. However, their abundance determinations were uncertain by factors of 2--3, since only one ion of each element was detected, leading to large and uncertain corrections for unobserved ions. 

In this work we combine optical and NIR spectroscopy to study NGC\,5315, a PN that has been classified as an almost spherical (slightly elliptical) PN with a complicated structure, including a somewhat broken ring.  The H-deficient WC4 central star has a temperature of about 76-79 kK \citep{marcolinoetal07,todtetal15}, which is considerably lower than other early-type [WR] stars (120$-$150 kK) . Several studies have focused on the chemical content in NGC\,5315, but draw different conclusions, especially regarding the N/O ratio which is important for constraining the progenitor mass. \citet{pottaschetal02} combined IUE ultraviolet and ISO infrared spectra with optical data to investigate chemical abundances, and found a relatively high N/O$\sim$0.88. \citet{milingoetal10} found a very similar value from optical/NIR spectrophotometry, as did \citet{peimbertetal04}.  The excess of He and N led these authors to classify NGC\,5315 as a Peimbert Type~I PN \citep[He/H~$>0.125$ and N/O~$>0.5$, or N/O~$>0.8$][]{peimbert78, kingsburghbarlow94}, although it is not an extreme object. The high concentration of He and N may be explained with the occurrence of the second dredge-up and CN-cycling during hot bottom burning (HBB), which are activated in AGB stars with M $>$ 3--4 M$_{\odot}$ \citep{beckeriben79, boothroydetal93, dicriscienzoetal16}. \citet{karakaslugaro16} found through detailed theoretical models that HBB activation requires a minimum progenitor mass between 4 and 5 M$_{\odot}$.

However, other investigations of NGC\,5315 \citep{defreitaspachecoetal91, samlandetal92, tsamisetal03, dufouretal15} do not find Type~I abundances, calling into question whether the PN indeed derives from a more massive progenitor.  It should be noted that only N$^+$ has collisionally-excited optical transitions.  Since this is a trace ion, the ionization correction factor (ICF) for N can be large and uncertain when only optical data are used in deriving its abundance.  The detection of \niii$]$ and \niv$]$ lines in UV spectra can lead to much more accurate N abundance determinations.  Both \citet{tsamisetal03} and \citet{dufouretal15} utilize optical and UV data, finding N/O ratios of 0.54 and 0.41, respectively, contrasting with the much higher value of \citet{pottaschetal02}. This variance may illustrate the uncertainties of abundances computed with highly temperature-sensitive UV lines.  As for the C/O abundance ratio, the values found in the literature range from 0.35 \citep{tsamisetal03} to 0.95 \citep{peimbertetal04}. The N/O and C/O ratios are very useful for constraining the progenitor mass, and the disparity in the ratios found, even from UV observations that provide access to a wide range of C and N ions, highlight the enigmatic nature of NGC\,5315's progenitor.  The different emission lines and methods used to derive C/O and N/O ratios will be discussed in detail in Sect.~\ref{sec:discuss}.

Abundance determinations of {\emph n}-capture elements in PNe can provide more restrictive constraints to the mass of the progenitor star. Observations of {\emph n}-capture elements in Type I PNe suggest that they exhibit little if any {\emph s}-process enrichments \citep{sterlingdinerstein08, sterlingetal15}, although in such studies only 1 or 2 {\emph n}-capture elements were studied. On the other hand, very low massive progenitors ($<1.5$~M$_{\odot}$) will also show little {\emph s}-process enrichments as they have not gone through the third dredge up. However Rb enrichments in PNe can help to disentangle this puzzle, because it is an indicator of the main neutron source during the AGB phase ($^{13}$C or $^{22}$Ne) and, hence of the mass of the progenitor star \citep{bussoetal99}.

There are relatively few observational constraints on nucleosynthetic models for AGB stars with  M $>$ 3--4 M$_{\odot}$. \citet{garciahernandezetal06} and \citet{garciahernandezetal09} reported large Rb enrichments in 4--8 M$_{\odot}$ Galactic and Magellanic cloud AGB stars, respectively. \citet{garciahernandezetal07} and \citet{garciahernandezetal13} provided Zr and Rb abundances, respectively, for massive galactic AGB stars; however, they found extremely high Rb/Zr ratios that were not predicted by nucleosynthesis models. \citet{zamoraetal14} took into account circumstellar effects on AGB stars to correct Rb abundance computations to Rb/Zr values in better agreement with results from AGB nucleosynthesis models for stars with masses between 4--8 M$_{\odot}$ (-0.2 $\le$ [Rb/Zr] $<$ 0.6) \citep{karakasetal12}. According to the models of \citet{karakasetal12}, Kr and Se should be enriched in the objects which show Rb enrichment for the $^{22}$Ne neutron source. Therefore, deep optical and NIR spectroscopy of PNe, with the aim of detecting emission lines of Rb, Kr and Se of different ionization stages, is a very valuable tool to improve the accuracy of these \emph{n}-capture element abundance determinations and hence, better constrain the initial mass of the progenitor stars of these objects.

This study is a continuation of our work to collect deep, high-resolution spectra of PNe to investigate objects covering different ionization degrees to detect as many ions as possible of {\emph n}-capture elements in order to improve the accuracy of their abundance determinations. These observations aim to address scientific goals including: i) to study the correlation between different {\emph n}-capture elements and C enrichment, predicted by models; ii) to study the correlation between the pattern of {\emph n}-capture element abundances and the mass of the progenitor star, which is modulated by the nuclear reaction activated in each mass range \citep{karakasetal12, vanraaietal12}; iii) to use detections of multiple ions of individual {\emph n}-capture elements to test the atomic data and ICF prescriptions \citep{sterlingetal15} for these species. Our first results for NGC\,3918 have been published by \citet{garciarojasetal15}. 

The observations and data reduction are described in Sect.~\ref{sec:obs}. The identification of lines and reddening correction is presented in Sect.~\ref{sec:id_lines}. In Sect.~\ref{sec:phys_cond} we compute the physical conditions. In Sects.~\ref{sec:ionic_ab} and~\ref{sec:total_ab} we compute ionic and total abundances. In Sect.~\ref{sec:discuss} we discuss the results and in Sect.~\ref{sec:conclusions} we draw some conclusions.

\begin{figure}
	\includegraphics[width=\columnwidth]{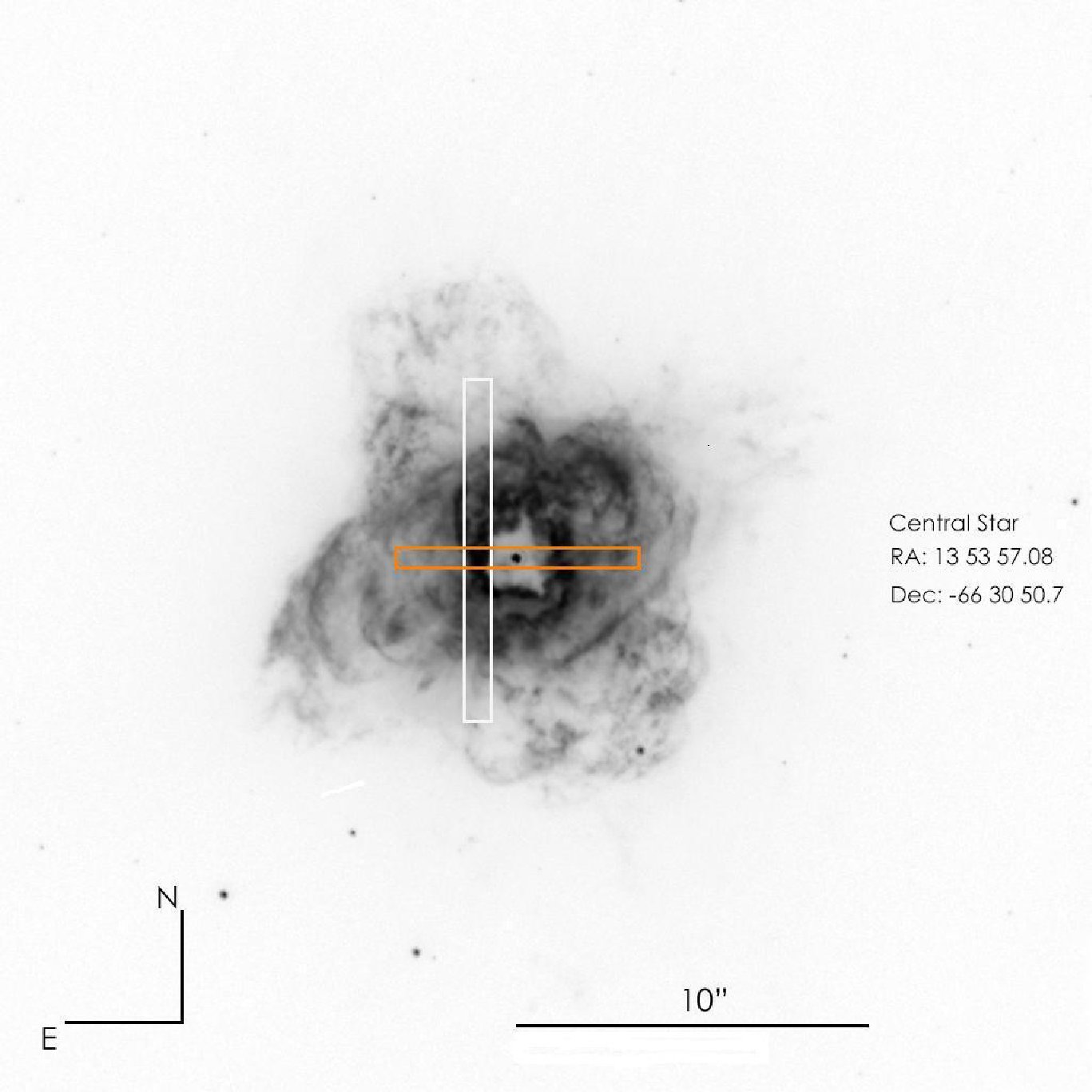}
    \caption{Deep Hubble Space Telescope (HST) H$\alpha$ image of NGC\,5315. Position and coordinate of the central star are shown. The N-S slit position for UVES observation is indicated as a white rectangular box, while the orange rectangular box represents the E-W slit for FIRE observation.}
    \label{slit}
\end{figure}

\begin{figure}
	\includegraphics[width=\columnwidth]{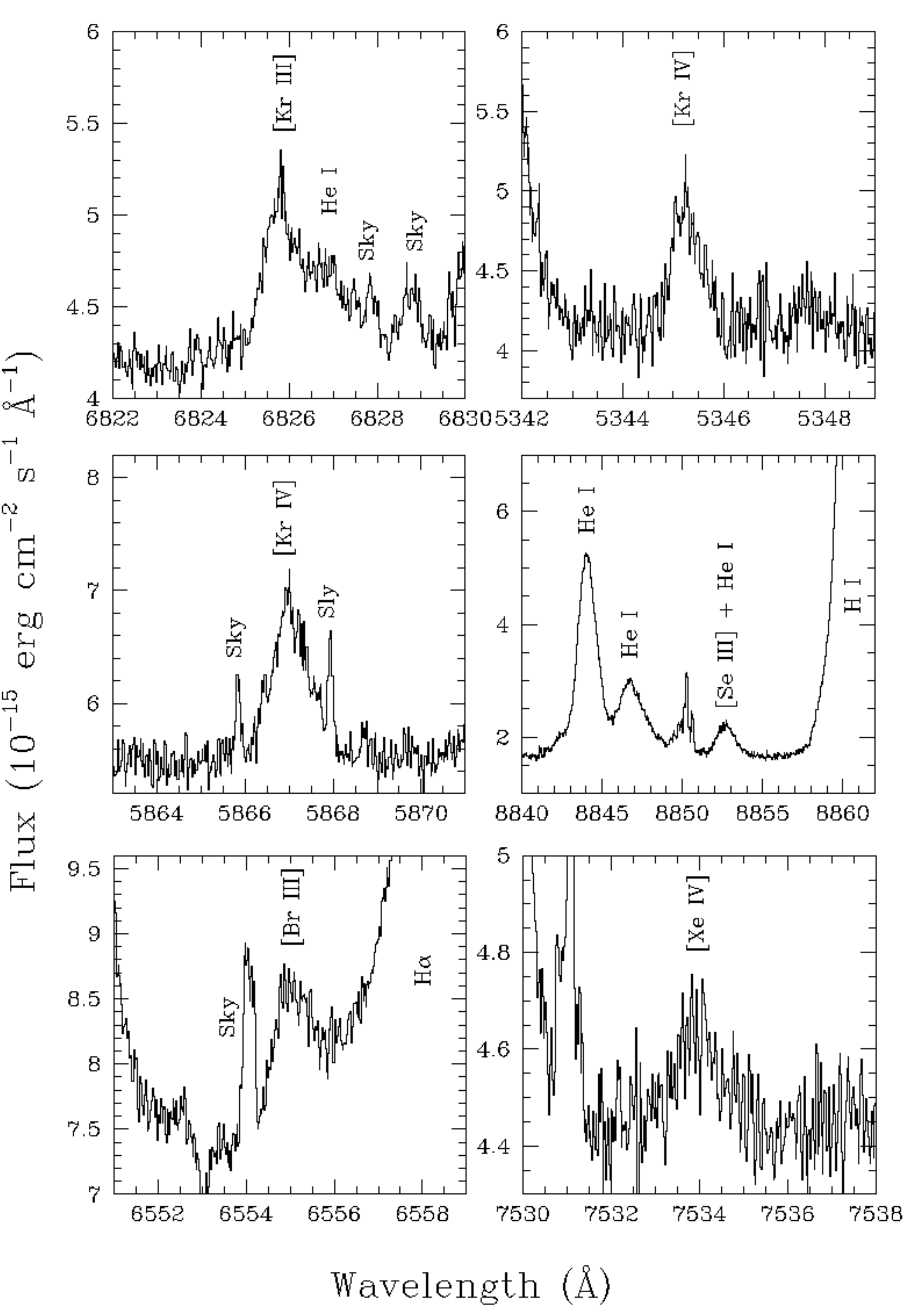}
	\includegraphics[width=\columnwidth]{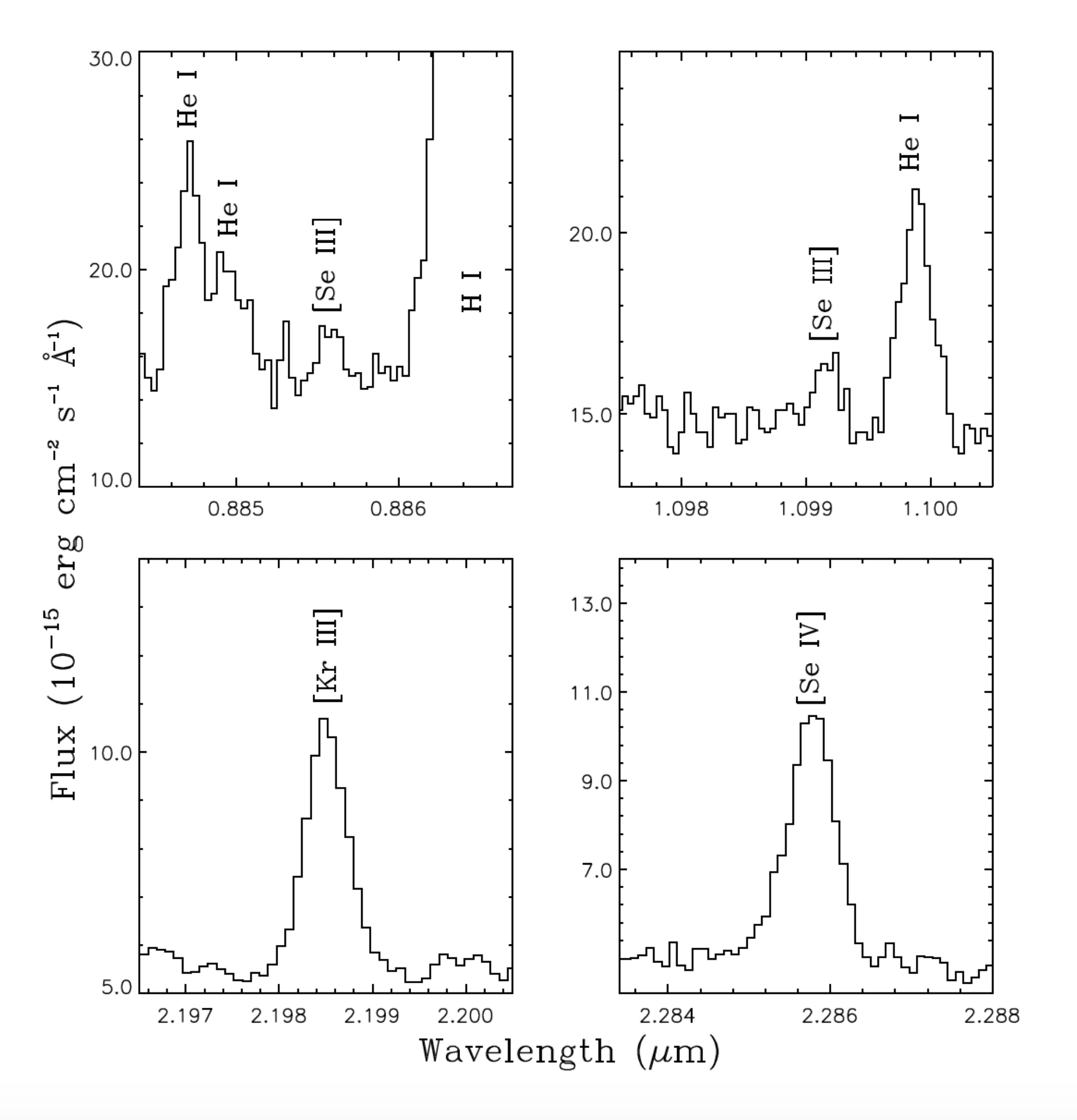}
    \caption{Upper panels: Line profiles of \emph{n}-capture element ions in the UVES spectrum, including the tentative detection of {\fbriii} $\lambda$6556 line. Lower panels: Detections of Se and Kr ions in the NIR FIRE spectrum.} 
    \label{s_profile}
\end{figure}

\section{Observations and data reduction}\label{sec:obs}

\subsection{UVES}
The optical spectra of NGC\,5315 were taken with the Ultraviolet-Visual Echelle Spectrograph \citep{dodoricoetal00}, attached to the 8.2m Kueyen (UT2) Very Large Telescope at Cerro Paranal Observatory (Chile) in service mode. The observations were performed during 3 nights under clear/dark conditions and the seeing remained below 1.5$''$ during the whole run (see Table~\ref{tobs}) . 

The slit width was set to 1 arcsec, which provides an effective spectral resolution \hbox{ R$\sim$ 40000}. See \citet{garciarojasetal15} for further details on the instrument setup.
The journal of observations is shown in Table~\ref{tobs}. The atmospheric dispersion corrector (ADC) was used to compensate for atmospheric dispersion. 
The UVES spectra are divided in four spectral ranges (B1, B2, R1 and R2). We took individual exposures of 630 s in each configuration and we followed the same sequence as in \citet{garciarojasetal15}. After combining all the extracted spectra we obtained a total exposure time of 2.10 h in each configuration. Additional single short exposures of 60 s during the first night were taken to obtain non-saturated flux measurements for the brightest emission lines. 
Reduction of the raw frames was done in the same way as in \citet{garciarojasetal15} for each night. The extracted spectra cover an area of 3.95$''$ $\times$ 1$''$ common to all spectral ranges.
Spectrophotometric standards Feige 67 \citep{oke90} and LTT\,3218 \citep{hamuyetal92, hamuyetal94} were observed to perform the flux calibration and were also fully reduced with the pipeline. We used {\sc iraf}\footnote{{\sc IRAF} is distributed by the National Optical Astronomy Observatories, which are operated by the Association of Universities for Research in Astronomy, Inc., under cooperative agreement with the National Science Foundation.} \citep{tody93} to perform the flux calibration and the radial velocity corrections. Flux-calibrated, radial velocity corrected, one-dimensional spectra for each night were finally co-added to obtain the final one-dimensional optical spectra analyses in this paper.

In Figure~\ref{slit} we show a high spatial resolution H$\alpha$ image of NGC\,5315 from the HST archive. The slit center was set 1.0$''$ east to the central star of NGC\,5315 oriented N-S (PA=0$^{\circ}$), covering the brightest area of NGC\,5315. 

\subsection{FIRE}
The NIR spectra of NGC\,5315 were taken with the Folded-port InfraRed Echellette \citep[FIRE, ][]{simcoeetal13} spectrograph attached to the 6.5m Magellan Baade Telescope (MBT) located at Las Campanas Observatory in Chile. The observations were performed on 2013 August 13 and the seeing oscillated between 1.8$''$--2.9$''$ during the run. 

The slit was located across the central star (PA=90$^{\circ}$) (see Figure~\ref{slit}) is 7$''$ long with a width of 0.75$''$, which lead to a resolution of R$\sim$ 4800. The spectral range is 0.8 -- 2.5 $\micron$, which covers J, H and K bands. Wavelength calibrations were performed in vacuum using a Th-Ar lamp, instead of the air wavelengths used in the UVES observations. Unfortunately, the lack of lines in the Th-Ar lamps at wavelengths larger than 2.3 $\mu$m led to a poor wavelength solution for the range 2.3-2.5 $\mu$m; however as this range is dominated by the Pfund {\hi} series, this did not affect the identification of the emission lines. For flux calibration and telluric correction, A0V standard stars were observed. We took 20 exposures of 30s, which led to a total integration time of 600s. We obtained 20 sky frames nodded 60$''$ along the slit direction, in order to subtract telluric features from the final spectra. Data reduction was performed with the FIRE reduction software package, FIREHOSE, using optimal extraction. This pipeline is based on the MASE pipeline for the MagE spectrograph \citep{bochanskietal09}, and a brief description of its performance can be found in \citet{simcoeetal13}. 
The journal of observations is shown in Table~\ref{tobs}.

\setcounter{table}{0}
\begin{table}
\begin{minipage}{75mm}
\centering
\caption{Journal of observations.}
\label{tobs}
\begin{tabular}{c@{\hspace{2.8mm}}c@{\hspace{2.8mm}}c@{\hspace{2.8mm}}c@{\hspace{1.8mm}}c@{\hspace{1.8mm}}}
\noalign{\hrule} \noalign{\vskip3pt}
Telescope& Date & $\Delta\lambda$~(\AA) & Exp. time (s) & Seeing ($''$) \\
\noalign{\vskip3pt} \noalign{\hrule} \noalign{\vskip3pt}
8.2 m VLT& 2014/01/02 & B1: 3100$-$3885 & 60, 2$\times$630 &$<$ 1.1 \\
"&" & B2: 3750$-$4995 & 60, 2$\times$630 & $<$ 1.1 \\
"&" & R1: 4785$-$6805 & 60, 2$\times$630 &  $<$ 1.1\\
"&" & R2: 6700$-$10420 & 60, 2$\times$630 &  $<$ 1.1\\
"& 2014/02/03 & B1: 3100$-$3885 & 6$\times$630 &$<$ 1.5 \\\
"&" & B2: 3750$-$4995 &  6$\times$630 & $<$ 1.5 \\
"&" & R1: 4785$-$6805 &  6$\times$630 &  $<$ 1.5\\
"&" & R2: 6700$-$10420 &  6$\times$630 &  $<$ 1.5\\
"& 2014/02/04 & B1: 3100$-$3885 & 4$\times$630 &$<$ 1.4 \\
"&" & B2: 3750$-$4995 &  4$\times$630 & $<$ 1.4 \\
"&" & R1: 4785$-$6805 &  4$\times$630 &  $<$ 1.4\\
"&" & R2: 6700$-$10420 &  4$\times$630 &  $<$ 1.4\\
6.5 m MBT & 2013/08/13 & 8000$-$25000 & 20$\times$30 &1.8--2.9 \\
\noalign{\vskip3pt} \noalign{\hrule} \noalign{\vskip3pt}
\end{tabular}
\end{minipage}
\end{table}

\section{Line fluxes, line identification and extinction correction}\label{sec:id_lines}

The line fluxes were measured with the \emph{splot} routine of the {\sc iraf} package. As in \citet{garciarojasetal15}, we decided to integrate over the entire line profile as set between two limits over a local continuum estimated by eye, as opposed to fitting an analytic function (such as a Gaussian) to the line profile. The UVES spectra present many misleading features such as telluric lines and internal reflections caused by the dichroic. For this reason, we had to take care to correctly distinguish nebular emission lines. Telluric emission lines are easily recognized and discarded, thanks to their peculiar shape in the 2D spectra (narrow profile and homogeneous emission along the slit).  For the weakest telluric features we used two catalogues: \citet{hanuschik03} in the optical and \citet{olivaetal15} in the NIR. When emission line intensities were affected by internal reflections or telluric absorption/emission, we include a note of caution in the line identification tables (Tables~\ref{lineid_uves} and~\ref{lineid_fire}). 

The UVES spectrum covers four spectral ranges, which overlap at the edges. In order to create a homogeneous set of data, we used the H9 $\lambda$3835 line, which lies in the overlapping region between the two spectral ranges B1 and B2, to normalize the line intensities, which were later re-scaled to {\hb}. Lines in the B2 and R1 ranges were directly scaled to {\hb} which is present in both ranges. {\fsii} $\lambda$6716 and $\lambda$6730 $\AA$ are the only lines common to the R1 and R2 settings. Unfortunately, in the R2 range, they lie at the extreme blue edge of the spectrum and their fluxes are not reliable. Therefore, we could not re-scale the fluxes in R2 to H$\beta$ and we normalized all the lines to P7 10047 $\AA$. The NIR FIRE spectrum does not include the {\hb} line, thus we used a different {\hi} line in each three spectral bands to normalize the line fluxes: P7 10047 $\AA$ for the J-band,  Br10 1.7367 $\mu$m for the H-band and Br$\gamma$ 2.1661 $\mu$m for the K-band. 

For the reddening correction, we assumed the extinction law of \citet{cardellietal89} with R$_{v}$=3.1. Using Balmer lines from UVES spectra we found a value of c(H$\beta$) = 0.63 $\pm$ 0.02. The deepest study of this object in the optical range until now is that of \citet{peimbertetal04}, which find a value of c(H$\beta$) = 0.74 $\pm$ 0.04; while other results from optical data give values in better agreement with our calculation: \citet{dufouretal15} find c(H$\beta$) = 0.56 from STIS/HST data and \citet{cahnetal92} compute c(H$\beta$) = 0.60. Since a normalization to H$\beta$ is not available for the spectral range R2 of UVES and for the J-band of the FIRE spectrum, we considered the extinction coefficient relative to P7 10047 $\AA$, c(P7)=0.05 $\pm$ 0.01, calculated assuming the extinction law by \citet{cardellietal89} with R$_{v}$=3.1. Finally, for the H-band and the K-band we used the theoretical ratios between the intensities of the reference lines used in each band (Br10 for the H-band and Br$\gamma$ for the K-band) and H$\beta$, provided by \citet{storeyhummer95} with {\elecd}=30,000 cm$^{-3}$ and {\te}=10,000 K, in order to re-scale the final intensities to H$\beta$.

We have detected about 700 emission lines. Many are permitted lines of {\hi}, {\hei}, {\oi}, {\oii}, {\nii}, {\ci}, {\cii}, {\siii}, {\nei}, {\neii}, {\mgii} and {\silii}. We also detect several forbidden and semi-forbidden lines from ions such as {\fnitroi}, {\fnii}, {\foi}, {\foii}, {\foiii}, {\fneiii}, {\sfmgi}, {\fpii}, {\fsii}, {\fsiii}, {\fclii}, {\fcliii}, {\fcliv}, {\fariii}, {\fariv}, {\fcrii}, {\fcriii}, {\fmniii}, {\ffeii}, {\ffeiii}, {\fkriii}, {\fkriv}, {\fseiii}, {\fseiv} and possibly {\fxeiv} and {\fbriii}. In the NIR spectrum, we also detect some molecular H$_2$ 1--0 transitions, but there are no signs of vibrationally excited H$_2$ lines. The depth of our spectra allows us to detect lines as faint as 10$^{-5}$$\times$\textit{I}(H$\beta$) and 10$^{-3}$$\times$\textit{I}({Br$\gamma$) in the optical and NIR spectra, respectively. The identifications and adopted laboratory wavelengths of the lines are based on several previous identifications in the literature \citep[e.~g.][and references therein]{garciarojasetal15, horalatterdeutsch99, peimbertetal04, rudyetal01}. We also made use of Peter van Hoof's atomic line list v2.05B18\footnote{http://www.pa.uky.edu/$\sim$peter/newpage/}. Details on the identification of \emph{n}-capture element emission lines are given in Sect.~\ref{sec:id}. 

To determine the line flux uncertainties, we considered individually each spectral range (B1, B2, R1, and R2) for UVES and the entire FIRE spectrum. Several lines were chosen in each of these ranges covering the whole range of measured fluxes, i.~e. 10$^{-5}\leq$ $F$($\lambda$)/$F$({\hb}) $\leq$10; uncertainties in individual fluxes were determined by choosing the highest and lowest reasonable values for the continuum by about 1$\sigma$ for the selected lines. To assign an error to the whole set of emission lines we made a logarithmic interpolation between relative intensities and measured uncertainties. An error of 5$\%$ for the flux calibration was added quadratically to all line flux uncertainties. In the final intensity uncertainties we also took in account the error for the extinction correction. Line intensities and identifications are presented in Table~\ref{lineid_uves} for the UVES spectrum and in Table~\ref{lineid_fire} for the FIRE spectrum. 

In order to deblend lines with almost coincident central wavelengths, we computed a model using {\sc Cloudy} v.13.03 \citep{ferlandetal13} to predicted line fluxes. Stellar temperature and luminosity, nebular diameter and elemental abundances of common elements were optimized to reproduce observed lines in the optical/NIR range from our data and select UV/IR lines from the \citet{pottaschetal02} spectra. The parameters obtained are very similar to those shown in Table 2 of \citet{sterlingetal15}. The model considers the whole PN, since we did not attempt to simulate an observation with a slit located on the PN. We believe this is a reasonable assumption owing to the fact that our slits cover all the ionization zones of the nebula. In Table~\ref{obs_pred} we compare different diagnostic ratios predicted by our photoionization model to the observed ones. In general the agreement between observed and predicted ratios is very good.

\setcounter{table}{3}
\begin{table}
	\centering
	\caption{Comparison of observed line ratios and those predicted by the {\sc Cloudy} model.}
	\label{obs_pred}
	\begin{tabular}{cccc} 
		\hline
		Ion & Line ratio & Model & Obs. \\
		\hline
		{\hei} & 6934/7161 & 0.437 & 0.448 \\
		{\hei} & 7161/7816    & 0.308 & 0.263 \\
		{\hei} & 10999/12988 & 0.426 &  0.492 \\
		{\fnii} & 5755/6538 & 0.0918 & 0.0816 \\
		{\foii} & 3726/3729 & 2.482 &  2.494 \\ 
		{\foiii} & 4363/5007 & 0.0048 &  0.0045 \\
		{\fsii} & 6731/6716 & 2.183 &  2.014 \\
		{\fsiii} & 6312/9069 & 0.0619 & 0.0571 \\
		{\fcliii} & 5538/5518 & 2.823 & 2.900 \\
		{\fariii} & 5192/7136 & 0.0057 &  0.0043 \\
		{\fariv} & 4740/4711 & 3.200  &  3.398 \\
		\hline
		
	\end{tabular}
\end{table}

\setcounter{table}{4}
\begin{table}
	\centering
	\caption{Corrected line ratios (\emph{I}(H$\beta$) = 100) for \emph{n}-capture elements in NGC\,5315.}
	\label{line_s}
	\begin{tabular}{cccc} 
		\hline
		$\lambda$$_{0}$ ($\AA$) & Ion & \emph{I}($\lambda$)/\emph{I}(H$\beta$) & Error(\%)$^{a}$\\
		\hline
		UVES\\
		5346.02 & {\fkriv} & 0.0081 & 20\\
		5867.74 & {\fkriv} & 0.0188 & 12\\
		6555.56 & {\fbriii} & 0.0097  & :\\
		6826.70 & {\fkriii} & 0.0161 & 16\\
		7535.40 & {\fxeiv} & 0.0032 & :\\
		8854.00 & {\fseiii} & 0.0055 & :\\
		\hline
		FIRE\\
		8855.28 & {\fseiii} & 0.0062 & :\\
		10992.00 & {\fseiii} & 0.0120 & :\\		
		2.1986 $\mu$m & {\fkriii} & 0.0556 & 13\\ 
                2.2864 $\mu$m & {\fseiv} & 0.0827 & 11\\ 
                \hline
                \multicolumn{4}{l}{${\rm ^a}$ Colons indicates errors larger than 40 $\%$}\
	\end{tabular}
\end{table}

\subsection{Identification of \emph{n}-capture ion lines}\label{sec:id}

As \emph{n}-capture elements have very low abundances \citep{asplundetal09}, their lines are very weak. This is why their detection in nebulae has always been a difficult task. However, since the pioneering work by \citet{pequignotbaluteau94}, this field has grown significantly. Several {\emph n}-capture element lines (particularly those of Kr, Xe and Se) have been detected in Galactic PNe and {\hii} regions, both in the optical \citep[e.~g. ][and references therein]{garciarojasetal15} and in the NIR \cite[e.~g.][]{dinerstein01, sterlingdinerstein08, blummcgregor08}, and even in other galaxies \citep{vanzietal08, mashburnetal16}.

We identify lines of the \emph{n}-capture element ions {\fkriii}, {\fkriv}, {\fseiii}, {\fseiv} and possibly {\fxeiv} and {\fbriii} in NGC\,5315 (see Figure~\ref{s_profile}). The measured fluxes are shown in Table~\ref{line_s}. Below, we give details on the identification of these lines in the UVES and FIRE spectra.

We identify lines of two ions of Kr. We detect the {\fkriii} 2.1986 $\mu$m line in the NIR spectrum. This line and the {\fseiv} 2.2864 $\micron$ line were first identified by \citet{dinerstein01}, and are the most widely detected \emph{n}-capture emission lines in PNe \citep{sterlingdinerstein08}. We also detect the faint {\fkriii} $\lambda$6826.70 line in our optical UVES spectrum. However, although our optical spectrum is of very high resolution, the velocity field of NGC\,5315 widens emission lines, and the flux of this line can be affected by  the faint {\hei} 3s $^{3}$S--16p $^{3}$P$^{0}$ $\lambda$6827.88 line \citep{pequignotbaluteau94}. We used the theoretical ratio between {\hei} lines of the same series (specifically, {\hei} $\lambda \lambda$6934, 7161 and 7816) using our {\sc Cloudy} model (see previous section) to estimate that this {\hei} line contributes 26\% of the total measured flux. The Kr$^{2+}$ abundance calculated with the corrected flux is in a good agreement with that from the {\fkriii} 2.1986 $\mu$m line (see Section~\ref{sec:ionic_ab_cels}). {\fkriii} $\lambda$9902.30 also falls in the spectral range of UVES and FIRE but is strongly blended with the relatively bright {\cii} $\lambda$9903.46 line. We find that the ratio {\fkriii} 9902.30/6826.70 = 0.078 for the assumed physical conditions (see Sect.~\ref{sec:phys_cond}), and hence {\fkriii} $\lambda$9902.30 accounts for only 1\% of the total measured flux of the line. Therefore we do not consider this {\fkriii} line to be detected. We also detect two {\fkriv} lines, $\lambda\lambda$5346.02 and 5867.74, which are the brightest \emph{n}-capture lines detectable in the optical spectra of PNe \citep[e.~g.][]{garciarojasetal15}. Given the relatively high brightness of these lines and the consistency between the computed abundances, we consider them well identified.

Thanks to the wide spectral range covered by our spectra, we detect three lines of Se ions. {\fseiii} $\lambda$8855.28 is detected in both UVES and FIRE spectra. It has almost the same wavelength as a weak He\,I $\lambda$8854.20 line, making its detection a delicate matter \citep{garciarojasetal15}. We used the theoretical ratios, computed with our {\sc Cloudy} model, of {\hei} lines belonging to the same series as {\hei} $\lambda$8854.20, to correct for the contribution of {\hei}. We find a contribution of 56$\%$ of {\fseiii} emission in the FIRE spectra and 53$\%$ in the UVES spectra, and the abundances calculated from both spectra are in good agreement (see Sect.~\ref{sec:ionic_ab_cels}). Nevertheless, we note that the line detected at 8855 $\AA$ is very faint and therefore the correction made with our photoionization model is very uncertain. Fortunately, we detect another line of Se$^{2+}$ at 10992 $\AA$ in the FIRE spectrum, which is isolated and was recently identified in NGC\,5315 by \citet{sterlingetal17}. We also detect the {\fseiv} 2.2864 $\mu$m line.

We possibly detect {\fxeiv} $\lambda$7535.40 line in our UVES optical spectrum. Unfortunately, the {\fxeiv} $\lambda$5709.21 line, which comes from the same upper level as {\fxeiv} $\lambda$7535.40, could not be observed since it is strongly blended with the permitted {\nii} $\lambda$5710.77 $\AA$ line. We computed {\fxeiv} $\lambda$5709.21/$\lambda$7535.40 = 0.82 for the assumed {\te} and {\elecd} (see Sect.~\ref{sec:phys_cond}). This corresponds to just 8\% of the 
flux measured for the feature at 5710 $\AA$ that we attribute to {\nii} $\lambda$5710.77 $\AA$, and thus we conclude that {\fxeiv} $\lambda$5709.21 line is not detected in our spectrum. Despite the non-detection of other Xe features in the spectrum of NGC\,5315, there are no reasonable alternative identifications to the 7535.40 line, and hence we identify it as {\fxeiv}.
On the other hand, owing to the ionization degree of NGC\,5315, Xe$^{2+}$ should be more abundant than Xe$^{3+}$. Therefore, we expect to detect the {\fxeiii} $\lambda$5846.77 line, which would strengthen the identification of the {\fxeiv} $\lambda$7535.40 line. Unfortunately, the {\fxeiii} $\lambda$5846.77 line lies on a continuum bump due to the extremely wide stellar {\civ} $\lambda$5808 line (the red bump of a [WC]-type star) which hampers any detection of faint lines. 

We also detect a feature that can be identified as {\fbriii} $\lambda$6555.56. However, we do not detect {\fbriii} $\lambda$6130.40 and therefore we cannot claim this as a unambiguous detection. 

We do not detect any emission line from Rb ions. We estimate an upper limit flux of {\frbiv} $\lambda$5759.55 line using a 3$\sigma$ criterion. We assume that the S/N in this region is too low to detect this line. Therefore, this estimation represents the maximum flux expected at that wavelength taking in account the continuum noise in that region. We also tried to estimate an upper limit to the {\frbiv} $\lambda$1.5973 $\mu$m line, but the telluric correction around this line was unreliable and an estimation of an upper limit is not possible. 

\setcounter{table}{5}
\begin{table}
\centering
\caption{Atomic data set used for collisionally excited lines.}
\label{atomic_cels}
\begin{tabular}{lcc}
\hline
& Transition  & Collisional \\
Ion & Probabilities & Strengths \\
\hline
N$^+$ & \citet{froesefischertachiev04} & \citet{tayal11} \\
O$^+$ & \citet{froesefischertachiev04} & \citet{kisieliusetal09} \\
O$^{2+}$ &  \citet{froesefischertachiev04} &  \citet{storeyetal14} \\
                &  \citet{storeyzeippen00} &  \\
Ne$^{2+}$ & \citet{galavisetal97} & \citet{mclaughlinbell00} \\
S$^+$ & \citet{podobedovaetal09} & \citet{tayalzatsarinny10} \\
S$^{2+}$ &  \citet{podobedovaetal09} & \citet{tayalgupta99} \\
Cl$^{+}$ & \citet{mendozazeippen83} & \citet{tayal04} \\
Cl$^{2+}$ & \citet{mendoza83} & \citet{butlerzeippen89} \\
Cl$^{3+}$ & \citet{kaufmansugar86} & \citet{galavisetal95} \\
         &  \cite{mendozazeippen82b} & \\
         &  \cite{ellismartinson84} & \\
Ar$^{2+}$ & \citet{mendoza83} & \citet{galavisetal95} \\
          &  \citet{kaufmansugar86} & \\
Ar$^{3+}$ & \citet{mendozazeippen82a} & \citet{ramsbottombell97} \\
Fe$^{2+}$ & \citet{quinet96} & \citet{zhang96} \\
          &  \citet{johanssonetal00} & \\
Se$^{2+}$ & \citet{sterlingetal17} & \citet{sterlingetal17} \\
Se$^{3+}$ & \citet{biemonthansen86a} & \citet{schoning97} \\
Kr$^{2+}$ & \citet{biemonthansen86a} & \citet{schoning97} \\
Kr$^{3+}$ & \citet{biemonthansen86b} & \citet{schoning97} \\
Br$^{2+}$ & \citet{biemonthansen86b} & \citet{schoning97}$^{a}$ \\
Xe$^{3+}$ & \citet{biemontetal95} & \citet{schoningbutler98} \\
\hline
\end{tabular}
\begin{description}
\item[$^{\rm a}$] Scaled from Kr$^{3+}$ effective collision strengths.
\end{description}
\end{table}

\section{Physical conditions}\label{sec:phys_cond}

Physical conditions were computed using flux ratios of ions which are sensitive to electron temperature and/or to electron density. The computations were carried out with PyNeb v1.0.26 \citep{luridianaetal15}, using the atomic data presented in Table~\ref{atomic_cels}. The electron density {\elecd} and temperature {\te} were estimated following the methodology described in previous works of our group \citep[e.~g.][]{garciarojasetal15}.  As in \citet{peimbertetal04}, we initially assumed three different zones in NGC\,5315, characterised by low (ionization potential IP $<$ 17 eV), medium (17 eV $<$ IP $<$ 39 eV) and high (IP $>$ 39 eV) ionization. However, given the relatively low excitation of NGC\,5315, and the similarities between different diagnostics, we adopted a two-zone model for {\elecd} (low-medium and high ionization zones) and {\te} (low and medium-high ionization zones) (see Table ~\ref{phy_cond}).

\begin{figure}
	\includegraphics[width=\columnwidth]{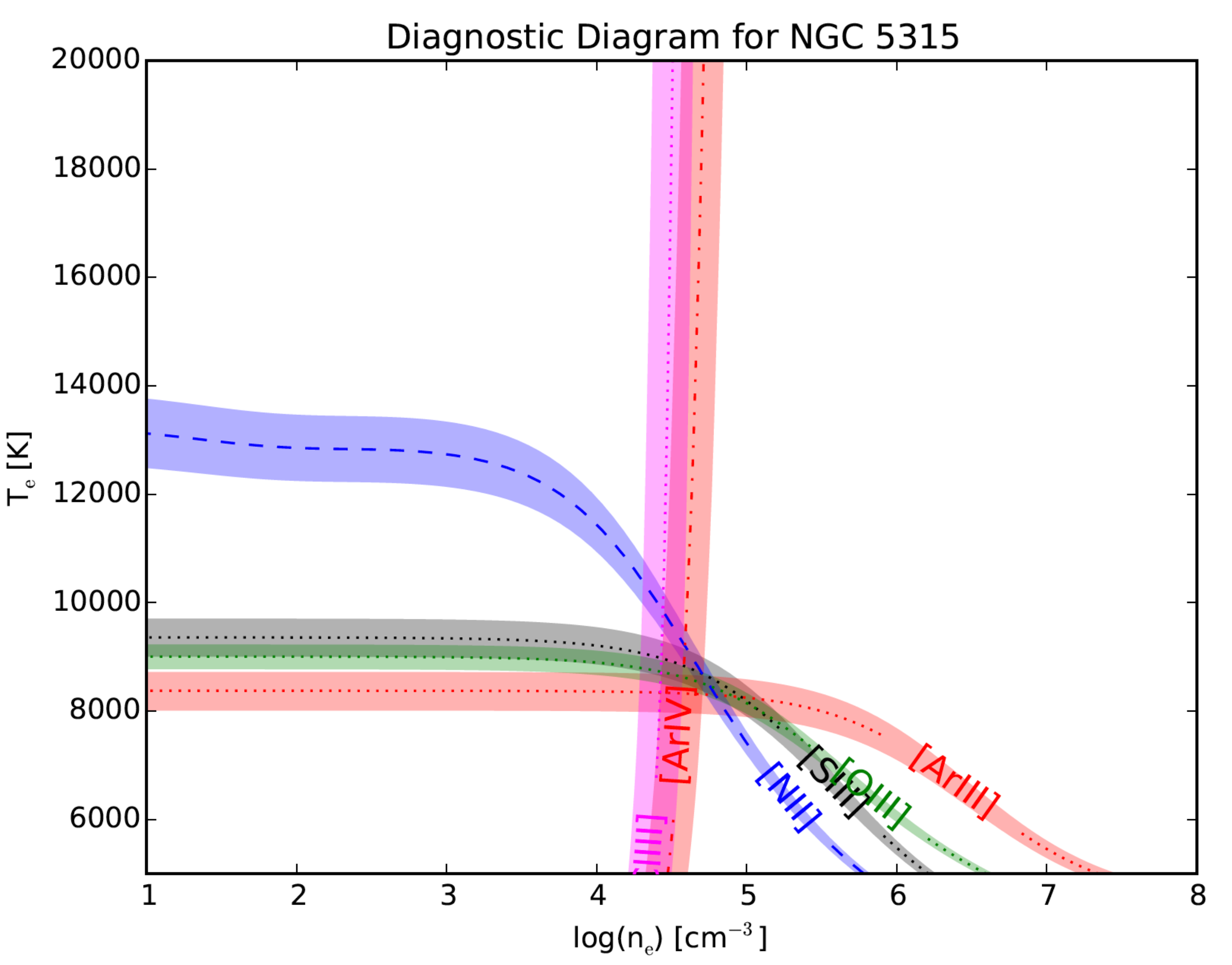}
    \caption{Diagnostic diagram for the complete set of ionization zones for NGC\,5315. Different colors are used for different elements, and different line styles correspond to different ions.}
    \label{diag}
\end{figure}

It is well known that the auroral {\fnii} $\lambda$5755 line is affected by recombination. We use Equation 1 of \citet{liuetal00} to compute the recombination contribution so that this line can be used for the {\te} computation. Since permitted {\nii} lines can be affected by fluorescence effects in relatively low ionization PNe \citep[see][]{escalanteetal12}, we made a preliminary computation of the N$^{2+}$/H$^+$ ratio by assuming N$^{2+}$/O$^{2+}$ (recombination lines) $\approx$ N$^{+}$/O$^{+}$ (collisionally excited lines). We find a recombination contribution os less than 5\% of the $\lambda$5755 flux, and therefore neglect it. We computed {\te} with the diagnostic {\fsiii} 6312/9069 using {\fsiii} $\lambda$6312 from UVES and  {\fsiii} $\lambda$9069 from FIRE data. Unfortunately, the {\fsiii} $\lambda$9069 line in the UVES spectrum is contaminated by atmospheric absorption and cannot be used. Moreover, the \fsiii} $\lambda$9530 line in the UVES spectrum is saturated and its measured flux is unreliable. As can be seen in Table~\ref{phy_cond}, {\te}({\fsiii}) is in good agreement with {\te}({\foiii}) and {\te}({\fariii}), calculated using lines from UVES data. A diagnostic diagram for NGC\,5315 is shown in Figure ~\ref{diag}.

Given the high density of NGC\,5315, the low ionization {\foii} $\lambda$3726/$\lambda$3729 and {\fsii} $\lambda$6731/$\lambda$6716 {\elecd} diagnostics are saturated, as it is clearly visible from the huge uncertainties of these ratios in Table~\ref{phy_cond}. Therefore, we decided to use the density found with the {\fcliii} $\lambda$5538/$\lambda$5518 diagnostic as representative of the low-medium ionization zone. 

We also calculated {\elecd} using {\ffeii} line ratios from our FIRE data, which are useful diagnostics for a large range of densities. Using PyNeb and the atomic dataset of Fe$^+$ by \citet{bautistaetal15}, we derive an electron density of \(30000\pm7000\) cm$^{-3}$ which is the average of the values from the {\ffeii} 1.2946 $\mu$m/1.2570 $\mu$m, 1.3281 $\mu$m/1.2570 $\mu$m, 1.5339 $\mu$m/1.2570 $\mu$m and 1.5999 $\mu$m/1.2570 $\mu$m diagnostics. This determination is consistent with the optical density diagnostics of the low-medium ionization zone (see Table~\ref{phy_cond}).

\setcounter{table}{6}
\begin{table}
	\centering
	\caption{Physical conditions in NGC\,5315.}
	\label{phy_cond}
	\begin{tabular}{llc} 
		\hline
		Diagnostic\\
		\hline
		\emph{n$_{e}$} (cm$^{-3}$)\\
		{\foii} $\lambda$3726/$\lambda$3729 &  &  30225$^{+25500}_{-7900}$\\
		{\fsii} $\lambda$6731/$\lambda$6716 &  &  9400:\\
		{\ffeii}$^{a}$ &  &  30000$\pm$7000\\
		{\fcliii} $\lambda$5538/$\lambda$5518 &  &  26800$^{+9000}_{-6500}$\\
		 \textbf{\emph{n$_{e}$}(low-mid) (adopted)} & &\textbf{26800$^{+9000}_{-6500}$} \\ 
                 {\fariv} $\lambda$4740/$\lambda$4711 &  &  38150$^{+12500}_{-9850}$\\
                 \textbf{\emph{n$_{e}$}(high) (adopted)} & &\textbf{38150$^{+12500}_{-9850}$} \\ 	
                 \\
                 \emph{T$_{e}$} (K)\\
                 {\fnii} $\lambda$5755/$\lambda$6584 &  &  9850$\pm$625\\
                 \textbf{\emph{T$_{e}$}(low) (adopted)} & &\textbf{9850$\pm$625}\\
                 {\foiii} $\lambda$4363/$\lambda$$\lambda$5007 &  &  8700$\pm$200\\
                 {\fariii} $\lambda$5192/$\lambda$7136 &  &  8350$\pm$350\\
                 {\fsiii} $\lambda$6312/$\lambda$9069 &  &  8975$\pm$350\\
                 \textbf{\emph{T$_{e}$}(mid-high) (adopted)} & &\textbf{8650$\pm$200}\\
                                                                                                                         \hline
                \multicolumn{3}{l}{${\rm ^a}$ Average value between different diagnostics (see text).}\\                                                                                                         
	\end{tabular}
\end{table}

The uncertainties in the physical conditions were calculated using Monte Carlo simulations. For each line flux involved in these calculations, we generated 10000 random values with a Gaussian distribution centered on the observed value with a sigma corresponding to the line flux error. The physical conditions with their uncertainties are presented in Table~\ref{phy_cond}.

\section{Ionic abundances}\label{sec:ionic_ab}

\subsection{Collisionally excited lines}\label{sec:ionic_ab_cels}

Ionic abundances were calculated using collisionally excited lines (CELs), for N, O, Ne, S, Cl, Ar, P, Fe, Se, Br, Kr and Xe ions. Based on the IP of the ions, we choose physical conditions of low ionization (IP $<$ 17 eV) for N$^{+}$, O$^{+}$, P$^{+}$, S$^{+}$, Cl$^{+}$, Fe$^{+}$,  and Fe$^{2+}$; medium ionization (17 eV $<$ IP $<$ 39 eV) for O$^{2+}$, S$^{2+}$,  Cl$^{2+}$, Cl$^{3+}$, Ar$^{2+}$, Se$^{2+}$, Se$^{3+}$, Br$^{2+}$ and Kr$^{3+}$; and high ionization (IP $>$ 39 eV) for Ne$^{2+}$ and Ar$^{2+}$. Computations were carried out using PyNeb and the atomic data used are presented in Table~\ref{atomic_cels}. Uncertainties in the line intensities and physical conditions were propagated via Monte Carlo simulations. Ionic abundances from CELs are presented in Table~\ref{ion_abu}.

To investigate the extent to which the different slit positions used for the UVES and FIRE observations affect our results, we compare the abundances of ions detected in both data sets (Cl$^{+}$, Cl$^{2+}$, P$^{+}$, Fe$^{+}$, Se$^{2+}$, and Kr$^{2+}$). As can be seen in Table~\ref{compare}, our optical ionic fractions from UVES spectrum are similar to both the FIRE and the \citet{peimbertetal04} data (our NIR observations were performed using a slit position similar to \citet{peimbertetal04}). In order to be consistent with our results, for this comparison we re-computed the ionic abundances from \citet{peimbertetal04} spectra, using our atomic data set. In Table~\ref{ion_abu} we can check that the abundances obtained from optical and NIR lines of Cl$^{+}$, Cl$^{2+}$, P$^{+}$, and Fe$^{2+}$ are in generally good agreement within uncertainties. 

For {\emph n}-capture element ions, we emphasize the good agreement between the abundances of Kr$^{2+}$ obtained from lines measured in the optical and in the NIR ranges, as well as the abundances obtained using the two available Kr$^{3+}$ lines in the optical. We also find good agreement between the Se$^{2+}$/H$^{+}$ abundances derived from {\fseiii} $\lambda$8855 in the UVES and FIRE spectra (12+log(Se$^{2+}$/H$^+$)=3.11 and 3.16, respectively).  The Se$^{2+}$ abundance derived from this line is $\sim$0.2 dex lower than that computed from the {\fseiii} $\lambda$1.0992 $\mu$m line, which we attribute to the uncertain correction for the contribution of \hei to the $\lambda$8854.20 flux.  We therefore choose to use the Se$^{2+}$ abundance obtained from the {\fseiii} $\lambda$1.0992 $\mu$m line in our analysis.

We also report the abundances of other 3 \emph{n}-capture element ions. We calculate the abundance of Br$^{2+}$ using the {\fbriii} $\lambda$6556 line and collision strengths scaled from Kr$^{3+}$ (see Table~\ref{atomic_cels}).  The abundance of Xe$^{3+}$ is calculated using the flux measured for the very faint {\fxeiv} $\lambda$7535.40 line. Finally, Rb$^{3+}$ is calculated from an upper limit flux estimation of the {\frbiv} $\lambda$5759.55 line. 

The FIRE NIR spectrum exhibits multiple {\ffeii} lines, which can be used as density diagnostics since they are not prone to fluorescent effects that contaminate optical {\ffeii} lines. 
The ratio of two lines arising from the same upper level is given by ratio of transition probabilities.
So, the expected {\ffeii} 1.2570 $\mu$m/1.6442 $\mu$m,  1.3209 $\mu$m/1.6442 $\mu$m, 1.3209 $\mu$m/1.2570 $\mu$m, 1.2946 $\mu$m/1.3281 $\mu$m, and 1.2946 $\mu$m/1.5339 $\mu$m line ratios are 1.30, 0.35, 0.26, 1.77 and 1.00, respectively, using the transition probabilities of \cite{bautistaetal15}.
The observed line ratios appear to be consistent, within the uncertainties, with the theoretical ones. 
In particular, the {\ffeii} 1.3209 $\mu$m/1.2570 $\mu$m, 1.2946 $\mu$m/1.3281 $\mu$m, and 1.2946 $\mu$m/1.5339 $\mu$m line ratios are 0.27$\pm$0.04, 1.70$\pm$0.40, and 0.83$\pm$0.16, respectively, which are in good agreement with theoretical expectations, whereas the {\ffeii} 1.2570 $\mu$m/1.6442 $\mu$m, and 1.3209 $\mu$m/1.6442 $\mu$m line ratios present values of 1.79$\pm$0.20 and 0.48$\pm$0.07, both higher than the theoretical ones. This is probably owing to an inaccurate subtraction of telluric emission nearby the {\ffeii} 1.6442 $\mu$m line that makes the flux measured for this line unreliable. In Table~\ref{ion_abu} we present the results obtained from individual {\ffeii} lines in both the UVES and FIRE wavelength ranges.

\setcounter{table}{7}
\begin{table}
	\centering
	\scriptsize
	\caption{Ionic abundances from CELs in units of 12 + log(X$^{\emph{i}+}$/H$^{+}$).}
	\label{ion_abu}
	\begin{tabular}{lllll} 
		\hline
		Ion &Line used  & Abu.$_{UVES}$ & &Abu.$_{FIRE}$ \\
		\hline
N$^{+}$	&	$\lambda$6548		&	7.67$\pm$0.08	\\
N$^{+}$	&	$\lambda$6584		&	7.70$\pm$0.08	\\
\textbf{N$^{+}$}& \textbf{(adopted)}   & &\textbf{7.69$\pm$0.08}  \\	
O$^{+}$	&	$\lambda$3726		&	7.96$^{+0.17}_{-0.15}$	\\
O$^{+}$	&	$\lambda$3729		&	7.96$^{+0.17}_{-0.15}$	\\
\textbf{O$^{+}$}& \textbf{(adopted)}   & &\textbf{7.96$\pm$0.17}   \\
O$^{2+}$	&	$\lambda$4959		&	8.70	$\pm$0.05	\\
O$^{2+}$	&	$\lambda$5007		&	8.72	$\pm$0.05	\\
\textbf{O$^{2+}$}& \textbf{(adopted)}   & &\textbf{8.71 $\pm$0.05} \\
Ne$^{2+}$	&	$\lambda$3868		&	8.11	$\pm$0.07	\\
Ne$^{2+}$	&	$\lambda$3968		&	8.11	$\pm$0.07	\\
\textbf{Ne$^{2+}$}& \textbf{(adopted)}   & &\textbf{8.11 $\pm$0.07} \\
P$^{+}$	&	$\lambda$7876		&	4.10	$\pm$0.12	\\
P$^{+}$	&	$\lambda$11470		&	&&4.09	$\pm$0.06	\\
P$^{+}$	&	$\lambda$11886&		&&4.28	$\pm$0.06\\
\textbf{P$^{+}$}& \textbf{(adopted)}  & &\textbf{4.24 $\pm$0.09}\\
S$^{+}$	&	$\lambda$6716		&	6.39$^{+0.12}_{-0.10}$	\\
S$^{+}$	&	$\lambda$6731		&	6.35$^{+0.12}_{-0.10}$	\\
\textbf{S$^{+}$}& \textbf{(adopted)}   & &\textbf{6.36$\pm$0.12}  \\
S$^{2+}$ & $\lambda$9069 &  &&7.08$\pm$0.03\\
S$^{2+}$ & $\lambda$9530 & &&7.13$\pm$0.03\\
\textbf{S$^{2+}$}	&	\textbf{(adopted)}		&	&\textbf{7.11	$\pm$0.03} \\
Cl$^{+}$	&	$\lambda$9123		&	4.33	$\pm$0.07 & \textbf{4.35$\pm$0.07} & 4.37 $\pm$0.07	\\
Cl$^{2+}$	&	$\lambda$3353		&	5.16	$^{+0.13}_{-0.11}$	\\
Cl$^{2+}$	&	$\lambda$5517		&	5.29	$\pm$0.08	\\
Cl$^{2+}$	&	$\lambda$5538		&	5.29	$\pm$0.05	\\
Cl$^{2+}$	&	$\lambda$8434		&	5.20	$^{+0.11}_{-0.09}$	\\
Cl$^{2+}$	&	$\lambda$8484	&		5.31	$^{+0.11}_{-0.09}$ & &5.42 $^{+0.12}_{-0.10}$	\\
\textbf{Cl$^{2+}$}& \textbf{(adopted)}   & & \textbf{5.29$\pm$0.07}\\
\textbf{Cl$^{3+}$}	&	\textbf{$\lambda$8046}			&\textbf{3.86	$\pm$0.06}	\\
Ar$^{2+}$	&	$\lambda$7135		&	6.61	$\pm$0.05	\\
Ar$^{2+}$	&	$\lambda$7751		&	6.62	$\pm$0.05	\\
\textbf{Ar$^{2+}$}& \textbf{(adopted)}   & &\textbf{6.61 $\pm$0.05}  \\
Ar$^{3+}$	&	$\lambda$4711		&	4.92	$\pm$0.10	\\
Ar$^{3+}$	&	$\lambda$4740		&	4.92	$\pm$0.06	\\
\textbf{Ar$^{3+}$}& \textbf{(adopted)}   & &\textbf{4.92 $\pm$0.08}  \\
Fe$^{+}$  &     $\lambda$7154   &    5.05$^{+0.12}_{-0.08}$ \\
Fe$^{+}$ &       $\lambda$8619 & & &5.12	$\pm$0.07	 \\
Fe$^{+}$ &      1.2570 $\mu$m & &&5.17	$\pm$0.06	\\
Fe$^{+}$ &      1.2946 $\mu$m & &&5.18	$\pm$0.07	\\
Fe$^{+}$ &      1.3209 $\mu$m & &&5.16	$\pm$0.07	\\
Fe$^{+}$ &      1.3281 $\mu$m & &&5.20	$\pm$0.09	\\
Fe$^{+}$ &      1.5339 $\mu$m & &&5.21	$\pm$0.07	\\
Fe$^{+}$ &      1.5999 $\mu$m & &&5.15	$\pm$0.09	\\
Fe$^{+}$ &      1.6642 $\mu$m & &&5.35	$\pm$0.11	 \\
Fe$^{+}$ &      \textbf{(adopted)} & &\textbf{5.15 $\pm$0.08} &	 \\
Fe$^{2+}$	&	$\lambda$4701		&	5.01	$\pm$0.10	\\
Fe$^{2+}$	&	$\lambda$4881		&	5.09	$\pm$0.10	\\
Fe$^{2+}$	&	$\lambda$5270		&	4.96	$\pm$0.11	\\
Fe$^{2+}$	&	2.2187 $\mu$m &	&&5.24	$\pm$0.15	\\
\textbf{Fe$^{2+}$}& \textbf{(adopted)}   & &\textbf{5.05 $\pm$0.12}\\	
Se$^{2+}$	&	$\lambda$8856	 	&3.11:	&&3.16:	\\
Se$^{2+}$	&	1.0995 $\mu$m		&	&&3.33	$\pm$0.16	\\
\textbf{Se$^{2+}$}& \textbf{(adopted)}   & &\textbf{3.33 $\pm$0.16}\\
\textbf{Se$^{3+}$}	&	\textbf{2.2864 $\mu$m}		&	&&\textbf{2.90	$\pm$0.05}	\\
Br$^{2+}$	&	$\lambda$6556		&	2.97:	\\	
Kr$^{2+}$	&	$\lambda$6826		&	3.34	$\pm$0.07 &	\\
Kr$^{2+}$	&	2.1986 $\mu$m	&		&  &3.39	$\pm$0.06 	\\
\textbf{Kr$^{2+}$}& \textbf{(adopted)}   & &\textbf{3.38 $\pm$0.06}&\\
Kr$^{3+}$	&	$\lambda$5346	&	3.00	$\pm$0.09	\\
Kr$^{3+}$	&	$\lambda$5867		&	3.12	$\pm$0.06	\\
\textbf{Kr$^{3+}$}& \textbf{(adopted)}   & &\textbf{3.08 $\pm$0.08} \\
\textbf{Rb$^{3+}$}	&	\textbf{$\lambda$5759}	&		\textbf{$<$2.80}	\\
\textbf{Xe$^{3+}$}	&	\textbf{$\lambda$7535}	&		\textbf{2.38:}	\\
                \hline
	\end{tabular}
\end{table}

\setcounter{table}{8}
\begin{table}
	\centering
	\caption{Comparison of ionic abundance ratios. The lines used to compute ionic abundances are listed in Table~\ref{ion_abu}.}
	\label{compare}
	\begin{tabular}{cccc} 
		\hline
		Ratio &UVES & FIRE & \citet{peimbertetal04}${\rm ^a}$\\
		\hline
O$^{+}$/O$^{2+}$ & 0.204$\pm$0.088& --& 0.153\\
Cl$^{3+}$/Cl$^{2+}$ & 0.037$\pm$0.009 & -- & 0.036\\
Cl$^{+}$/Cl$^{2+}$& 0.110$\pm$0.028& 0.089$\pm$0.032& 0.089\\
Ar$^{3+}$/Ar$^{2+}$ & 0.020$\pm$0.005 &-- & 0.011\\	
Fe$^{+}$/Fe$^{2+}$& 1.041$\pm$0.442 &  0.844$\pm$0.388 & --\\
                \hline
\multicolumn{3}{l}{${\rm ^a}$ Computed with atomic data shown in Table~\ref{atomic_cels}}\\
	\end{tabular}
\end{table}

\subsection{Optical recombination lines}\label{sec:ionic_ab_rls}

\begin{figure*}
	\includegraphics[width=\textwidth]{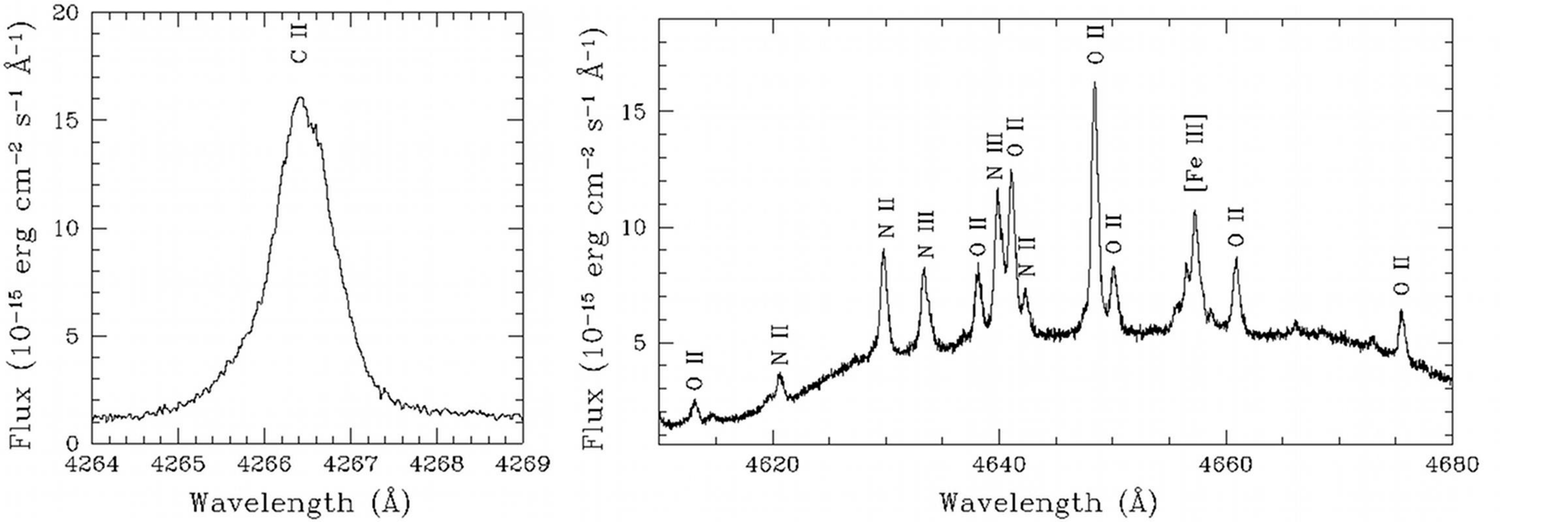}
    \caption{Portion of the echelle optical spectra of NGC\,5315, showing the zone where the {\cii} $\lambda$4267 line and the multiplet 1 {\oii} lines lie. A bump in the continuum in the right plot is apparent, and arises from a strong stellar feature, which is a blend of stellar {\ciii}, {\civ} and {\heii} broad emission lines \citep{marcolinoetal07}.}
    \label{cii_oii_orls}
\end{figure*}

We calculate ionic abundances using He, C, N, O and Ne optical recombination lines (ORLs). The atomic data used are shown in Table~\ref{atomic_rls}. We detect several permitted lines of heavy-element ions, such as {\cii}, {\nii}, {\niii}, {\oi}, {\oii}, {\oiii}, {\neii} and {\mgii}, but some are affected by fluorescence or contaminated by telluric features. We only consider pure recombination lines in these calculations. Discussion about the formation mechanism of several permitted lines can be found in \citet{estebanetal98} and \citet{estebanetal04}. In Fig.~\ref{cii_oii_orls} we show the brightest {\cii} and {\oii} lines in our optical spectrum. The high resolution of the data allow to deblend the multiple features in the region of the multiplet 1 {\oii} lines.

The abundance discrepancy problem arises from the fact that ORLs provide ionic abundances that are systematically larger than those obtained from CELs in photoionized nebulae, whether {\hi}i regions or PNe.  Solving this problem has critical implications for the measurement of the chemical content of nearby and distant galaxies, which is most often done using CELs from their ionized interstellar medium. Three main scenarios have been proposed to explain this discrepancy: i) the existence of temperature fluctuations over the observed volume of the nebula \citep{peimbert67, torrespeimbertetal80}; b) cold and dense H-poor inclusions in which the bulk of the ORL emission originates \citep[e.~g.][]{liuetal00} and, c) the departure of the free electron energy distribution from the Maxwellian distribution \citep[$\kappa$-distribution, see][]{nichollsetal12}. While there is no direct observational evidence that favours or discards any of these scenarios, indirect evidence suggests that metal-rich inclusions may be the source of ADFs in PNe \citep{tsamisetal04, wangliu07, garciarojasetal16}.

Importantly, previous studies have shown that the ADFs of C, N, and O ions in Galactic PNe are very similar \citep{liuetal00, wangliu07}, and hence abundance ratios found from ORLs are suitable proxies for CEL abundance ratios.  This is most relevant in the case of C/O, since no collisionally-excited lines of C ions lie in the UVES/FIRE spectral range.

NGC\,5315 has a relatively low degree of ionization, and thus we do not detect {\heii} lines. He$^{+}$ abundances are computed with PyNeb, using the three brightest {\hei} lines $\lambda\lambda$4471, 5876 and 6678. The effects of collisional contributions and optical depth in the triplet lines are considered. Results from the three lines are very consistent and are shown in Table~\ref{abu_rel}.

C$^{2+}$ abundances from ORLs are in very good agreement between lines belonging to different multiplets (see Table~\ref{abu_rel}). {\cii} $\lambda$9903.46 is blended with the very faint {\fkriii} $\lambda$9902.30 line. Although this contribution should be negligible, we do not consider this line, because it yields a higher abundance from FIRE data than from UVES data. We ascribe this discrepancy to atmospheric absorption effects visible in the 2D optical spectrum. In general, C$^{2+}$ abundances from ORLs are in very good agreement with those computed by \citet{peimbertetal04} for this object.

\citet{fangetal11} compute very detailed recombination coefficients for {\nii} lines. We have used these atomic data to compute N$^{2+}$ abundances from multiplet 3 {\nii} lines. We find an excellent agreement between all the lines of the multiplet.

To compute O$^{2+}$ abundances from ORLs we use lines from multiplets 1, 2, 10 and 20 as recommended by \citet{estebanetal04}. Given the high density of NGC\,5315, departures from LTE for {\oii} multiplet 1 \citep{ruizetal03, tsamisetal04} that are important when {\elecd} $<$ 10$^{4}$ cm$^{-3}$ can be ignored. The agreement in O$^{2+}$ abundances derived from the different multiplet 1 lines shows that this is a good assumption (Table~\ref{abu_rel}).  We use recombination coefficients assuming LS-coupling from \citet{storey94} for 3s-3p transitions (multiplets 1 and 2), and the intermediate-coupling scheme by \citet{liuetal95} for 3p-3d transitions (multiplets 10 and 20). Results are shown in Table~\ref{abu_rel}.

We compute Ne$^{2+}$ abundances from ORLs using lines of multiplets 1, 39, 55 and 57. A detailed discussion of the excitation mechanisms of these lines can be found in \citet{garciarojasetal15}. We use recombination coefficients by \citet{kisieliusetal98} for multiplets 1 and 39 and those by Kisielius (private communication) for multiplets 55 and 57. We adopted the average abundance given from all these multiplets. 

Finally, we estimate Mg$^{2+}$ abundance from the {\mgii} 3d--4f $\lambda$4481 line. For this ion there are no recombination coefficients available, so we assume that the {\mgii} $\lambda$4481 line has an effective recombination coefficient equal to that of the {\cii} $\lambda$4267 line \citep{barlowetal03, wangliu07}.

\setcounter{table}{9}
\begin{table}
\centering
\begin{minipage}{180mm}
\caption{Atomic data set used for recombination lines.}
\label{atomic_rls}
\begin{tabular}{lc}
\hline
Ion & Recombination Coefficients \\
\hline
H$^{+}$ &  \citet{storeyhummer95}  \\
He$^{+}$ &  \citet{porteretal12, porteretal13}  \\
C$^{2+}$ & \citet{daveyetal00}  \\
N$^{2+}$ & \citet{fangetal11, fangetal13}  \\
O$^{2+}$ &  \citet{storey94}  \\
                &  \citet{liuetal95}  \\
Ne$^{2+}$ &  \citet{kisieliusetal98}  \\
                &  Kisielius \& Storey (private communication) \\
Mg$^{2+}$ &  \citet{kisieliusetal98}  \\
\hline
\end{tabular}
\end{minipage}
\end{table}

\setcounter{table}{10}
\begin{table}
        \centering
        \caption{Ionic abundances from ORLs.}
        \label{abu_rel}
        \begin{tabular}{lcc} 
                \hline
                Mult. & $\lambda$$_{0}$ ($\AA$)& X$^{i+}$/H$^{+}$(10$^{-5}$)\\
                \hline
                & He$^{+}$ & \\
                14 & 4471 & 12267$\pm$860\\
                11 & 5876 & 13150$\pm$924\\
                46 & 6678 & 12219$\pm$853\\
                & \textbf{adopted} & \textbf{12819$\pm$880}\\
                \hline
                & C$^{2+}$ & \\
                6 & 4267.15 & $\textbf{66$\pm$5}$\\
                16.04 & 6151.43 & $\textbf{61$\pm$6}$\\
                17.02${\rm ^a}$ & 9903.46 & 77$\pm$6\\
                17.06 & 5342.38 & $\textbf{62$\pm$6}$\\
                & \textbf{adopted} & \textbf{63$\pm$6}\\
                \hline
                & N$^{2+}$ &\\
                3 & 5666.64 & 61$\pm$6\\
                 & 5676.02 & 48$\pm$7 \\
                 & 5679.56 & 49$\pm$5\\
                 & 5686.21 & 53$\pm$8\\
                 & 5710.76 & 51$\pm$8\\
                 & 5730.65 & 69:\\
                &  \textbf{adopted} & \textbf{52$\pm$4}\\
                \hline
                & O$^{2+}$ &\\
                1 & 4638.85 & 89$\pm$11\\
                 & 4641.81 & 80$\pm$7\\
                 & 4649.14 & 81$\pm$6\\
                 & 4650.84 & 87$\pm$11\\
                 & 4661.64 & 83$\pm$9\\
                 & 4676.24 & 61$\pm$9\\
                & sum & \textbf{81$\pm$7}\\
                2${\rm ^b}$ & 4317.14$^{d}$ & 84$\pm$18\\
                 & 4319.63 & 59$\pm$14\\
                 & 4325.76$^{c}$ & 242$\pm$73\\
                 & 4336.83 & 76$\pm$28\\
                 & 4345.56 & 103$\pm$19\\
                 & 4349.43 & 65$\pm$9\\
                 & 4366.89 & 116$\pm$19\\
                & sum$^{e}$ & \textbf{81$\pm$17}\\
                10$^{b}$ & 4072.15 & \textbf{81$\pm$9}\\
                 & 4092.93 & 66:\\
                20 & 4110.79 & 121$\pm$8\\
                 & 4019.22 & 59$\pm$12\\
                & average & \textbf{81$\pm$12}\\
                \\
                & \textbf{adopted} & \textbf{81$\pm$12}\\
                \hline
                & Ne$^{2+}$ &\\
                1 & 3694.22 & 21$\pm$6\\
                 & 3709.62 & 27$\pm$9\\
                 & 3766.26 & 25$\pm$10\\
                 & 3777.14 & 27$\pm$11\\
                & sum & \textbf{24$\pm$6}\\
                39 & 3829.77 & 18:\\
                55 & 4409.30 & \textbf{11$\pm$4}\\
                57 & 4428.77 & 4:\\
                & \textbf{adopted} & \textbf{19$\pm$6}\\
                \hline
                & Mg$^{2+}$ &\\
                1 & 4481.21 & 3.6$\pm$0.8\\
                                                  \hline
                \multicolumn{3}{l}{$^{\rm a}$ Blended with [Kr III] $\lambda$9902.30}\\
                  \multicolumn{3}{l}{$^{\rm b}$ Assuming case A.}\\
                   \multicolumn{3}{l}{$^{\rm c}$ Blended (see table~\ref{lineid_uves}).}\\
                   \multicolumn{3}{l}{$^{\rm d}$ Affected by charge transfer.}\\
                   \multicolumn{3}{l}{$^{\rm e}$ Without considering blended lines.}\\
        \end{tabular}
\end{table}

\section{Total abundances}\label{sec:total_ab}

To compute total abundances we have to take unobserved ions into account. 
In general for the most common elements we used the detailed ionization correction factors (ICFs) provided by \citet[][hereinafter D-I14]{delgadoingladaetal14} to correct for the presence of unobserved ions. However, we make an exception for N (see below). 
In Table~\ref{abu_tot} we present the total abundances we computed for NGC\,5315. The first 4 columns of Table~\ref{abu_tot} present the abundances obtained from our data from both CELs and ORLs, either by summing the ionic abundances or by applying an ICF.  The next columns show the results obtained by \citet{peimbertetal04}, \citet{pottaschetal02}, \citet{tsamisetal03} and \citet{tsamisetal04}. The last column of Table~\ref{abu_tot} shows the abundances from our optimized {\sc Cloudy} model. As can be seen in Table~\ref{abu_tot}, there is general good agreement within the uncertainties between our computed elemental abundances and those derived by our model and in previous works. 
Below we briefly discuss the most common elements, and in Section~\ref{sec:n_capt_abu} we discuss elemental abundances of {\emph n}-capture elements. 

In the spectrum of NGC\,5315 there are no {\heii} lines and ions with similar or higher IP are not seen. Given that \hbox{(O$^{2+}$/(O$^{+}$+O$^{2+}$) = 0.85}, the contribution of neutral He to the total He abundance should be negligible (D-I14) and thus we assume that He/H = He$^{+}$/H$^{+}$. 

Similarly, we can disregard the presence of O$^{3+}$ in the nebula. Thus, we computed O abundance from CELs by simply adding O$^{+}$ and O$^{2+}$ abundances, which are the only two ions of this element detected in the spectrum of NGC\,5315. We estimated O$^{+}$ from ORLs by scaling the ratio O$^{+}$/O$^{2+}$ computed from CELs. The total O abundance from ORLs is higher, resulting in an abundance discrepancy factor, ADF(O)=1.58 (see Table~\ref{abu_tot}).

\setcounter{table}{11}
\begin{table*}
 \caption{Total abundances (12 + log(X/H)).}
 \label{abu_tot}
 \begin{tabular}{lccccccccccccc}
  \hline
 &  &  &  &  & \multicolumn{2}{c}{PE04$^a$} & & \multicolumn{2}{c}{PO02$^b$} & & \multicolumn{2}{c}{T03, T04$^{c}$}  & \\
\cline{6-7}
\cline{9-10}
\cline{12-13}
  Element & CELs sum & CELs ICF & ORLs sum & ORLs ICF & CELs  & ORLs & & CELs&  ORLs& & CELs& ORLs & Model \\
  \hline
  He & -- & -- & 11.11$\pm$0.03 & -- &  -- & 11.09 & & -- & 11.09 & & -- & 11.08 & 11.03\\
  C   & -- & -- & 8.83 $\pm$0.05$^{h}$ & 8.89$\pm$0.10& -- & 8.85 & & 8.64 & -- & & 8.33 &8.86 & 8.42\\
  N &--& 8.52$^{+0.30}_{-0.21}$$^{d}$ & -- & 8.79$^{+0.16}_{-0.08}$$^{\rm i}$& -- & 8.82 & & 8.66  &-- & &8.52 &8.77 & 8.39 \\
  O & 8.78$\pm$0.05 & -- &8.98$\pm$0.06 &--& 8.63 & 8.87& & 8.72  &--  & & 8.79  &9.10 & 8.80\\
  Ne &--&8.40$\pm$0.12 & -- & -- & 8.05 &-- & & 8.20 &-- & & 8.30 &-- & 8.09\\
  Mg & -- & --  & 7.56$\pm$0.10 & -- & -- & -- & & -- &--  & & -- &-- & -- \\
  P & -- & 5.06:  &-- & -- & -- & -- & & -- &--  & & -- &-- & 5.30\\
  S & 7.18$\pm$0.03 & -- & -- & -- & 7.34 & -- & & 7.08 & -- & &7.31 & -- & 7.27\\
  Cl & 5.35$\pm$0.07 & -- & -- & -- & 5.36 & -- & & -- & -- & & 5.41  &-- & 5.37\\
  Ar & -- & 6.73$\pm$0.22 &-- & -- & 6.56  &-- & & 6.66  &-- & & 6.56  &-- & 6.49\\
  Fe & 5.40$\pm$0.10 & 5.77 $\pm$0.22/5.48$\pm$0.17$^{g}$ &-- & -- & --  &-- & & --  &-- & & --  &-- & 5.70 \\
  Se & -- & 3.60$\pm$0.17$^{e}$& -- & -- & -- & -- & & -- & -- & & -- & -- & -- \\
  Kr & -- & 3.60$\pm$0.07$^{f}$ & -- & -- & -- & -- & & -- & -- & & -- & -- &  --\\
  Xe & -- & 3.43:& -- & -- & -- & -- & & -- & -- & & -- & -- & -- \\
  Br & -- & 3.53: & -- & -- & -- & -- & & -- & -- & & -- & -- & -- \\
  Rb & -- & $<$2.87 & -- & -- & -- & -- & & -- & -- & & -- & -- & -- \\
      \hline
   \multicolumn{14}{l}{$^a$ \citet{peimbertetal04}}\\
   \multicolumn{14}{l}{$^b$ \citet{pottaschetal02}}\\
    \multicolumn{14}{l}{$^c$ \citet{tsamisetal03, tsamisetal04}}\\
  \multicolumn{14}{l}{$^d$ Using the ICF N/O=N$^{+}$/O$^{+}$}\\
  \multicolumn{14}{l}{$^e$ Using the ICFs given by eq. 9 of \citet{sterlingetal15}}\\
  \multicolumn{14}{l}{$^f$ Using the ICF given by eq. 4 of \citet{sterlingetal15}}\\
  \multicolumn{14}{l}{$^g$ Using the ICFs of eq. 2/eq. 3 provided by \citet{rodriguezrubin05}}\\
    \multicolumn{14}{l}{$^h$ Scaling C$^{2+}$/C$^{+}$ from CELs using an average value between \citet{dufouretal15} and \citet{pottaschetal02} data}\\
    \multicolumn{14}{l}{$^i$ Assuming N/O~=~N$^{2+}$/O$^{2+}$ (ORLs)}\\
  \end{tabular}
\end{table*}

Although C CELs are only detectable in UV spectra, C abundances can also be computed from optical {\cii} and {\ciii} ORLs. However, the excitation of NGC\,5315 is too low to detect {\ciii} RLs. We compute the elemental C abundance using our C$^{2+}$ abundance from {\cii} ORLs and the ICF presented in equation 39 of D-I14. We compare our results with that obtained assuming a simple sum of C$^{+}$ and C$^{2+}$ abundances from ORLs; to estimate C$^{+}$/H$^{+}$ we rescale using the C$^{+}$/C$^{2+}$ ratios obtained from CELs by \citet{pottaschetal02} and by \citet{dufouretal15}. Both results are in very good agreement with what we found using the ICF correction (see table~\ref{abu_tot}). Despite the well-known abundance discrepancy problem between CEL and ORL abundances (see Sect.~\ref{sec:ionic_ab_rls}), numerous studies have shown that C/O (and N/O) ratios computed from ORLs are consistent with the ratios derived from CELs \citep[e.g.,][]{tsamisetal04, wessonetal05, wangliu07, delgadoingladarodriguez14}.

To compute the N elemental abundance from optical and NIR spectra is a delicate matter. We have computed N/O using all the available lines and different methods to better constrain this abundance ratio.  First, we used the {\fnii} lines to compute N$^+$/H$^+$. D-I14 argue that the classical scheme N/O = N$^+$/O$^+$ may underestimate N abundances, especially when the temperature of the central star is low, which is the case when {\heii} lines are not observed, and they proposed a new ICF. However, in a later work, \citet{delgadoingladaetal15} found that their ICF could introduce an unexpected trend with the degree of ionization in the N/O values obtained for a group of {\hii} regions and PNe. The trend seems to be related to the fact that the nebulae are either matter- or density-bounded.  The classical ICF seems to produce more accurate results for radiation-bounded nebulae, while the ICF from D-I14 is preferred for matter-bounded nebulae. To better constrain the N ICF, we compared observed ionic fractions (O$^{2+}$/O$^{+}$, S$^{2+}$/S$^{+}$, Ar$^{+3}$/Ar$^{2+}$, and He$^{2+}$/He$^{+}$) to the grid of photoionization models of D-I14. From our inspection we concluded that NGC\,5315 is most likely radiation-bounded, and hence the classical ICF provides a more reliable value of the N/O in this PN.  The derived N/O value from CELs and the classical ICF is 0.55.  The uncertainties in log(N/O) associated with this ICF, estimated from the grid of photoionization models by D-I14, are $^{+0.3}_{-0.2}$ dex.

Secondly, our optimized Cloudy model (Section~\ref{sec:id_lines}) produces a somewhat lower N abundance than the one derived with the ICF method (12+log(N/H)=8.39 compared to 8.52), and so is the N/O ratio (0.39). As a third estimate, we computed the total N/O ratio from ORLs (see Table~\ref{abu_tot}).  Since permitted optical {\nitroi} lines are strongly affected by fluorescence effects, it is not possible to determine the N$^+$/H$^+$ abundance from ORLs.  Therefore we use the approximation N/O~=~N$^{2+}$/O$^{2+}$ for ORLs, which gives a ratio of 0.64. The uncertainties in log(N/O) associated with this ICF are $^{+0.15}_{-0.05}$ and they were obtained from the D-I14 grid of photoionization models. We use the average N/O ratio weighted by the uncertainties from these estimates, 0.59$^{0.24}_{0.15}$, in the remainder of our analysis.

We calculate total abundances of Ne, S, Cl and Ar, using the ICFs developed by D-I14 when needed, or by summing the ionic abundances when a sufficient number of ions are detected.
The optical range hosts several {\fneiii}, {\fneiv}, and {\fnev} CELs, but owing to the relatively low ionization degree of NGC\,5315 we only detect {\fneiii} lines. There are no lines of the Ne$^{+}$ ionization stage in our spectral range, and to correct for its contribution we used the ICF given by equation 17 of D-I14.  We detect {\fsii} and {\fsiii} lines in our spectra. D-I14 found that if {\fsii} and {\fsiii} are seen in the spectra and the object does not show {\heii} emission, more highly-charged states have a negligible contribution. Thus, we compute the total S abundance by adding the S$^+$ and S$^{2+}$ ionic abundances. Our results are consistent with values from literature (see Table~\ref{abu_tot}). Cl ions have several emission lines in the optical range, including {\fclii}, {\fcliii} and {\fcliv}. The amount of Cl$^{4+}$ is negligible for NGC\,5315, because the IP is too high (53.5 eV). We follow the recommendation by D-I14, who suggest that total Cl abundance can be computed as the sum of ionic abundances when {\fclii}, {\fcliii} and {\fcliv} lines are detected and {\heii} lines are not seen in the spectra. Finally, we detect {\fariii} and {\fariv} lines. Given the ionization degree of NGC\,5315, we have to take into account the contribution of Ar$^{+}$, which can be an important contributor to the total abundance of Ar in low ionization PNe. The contribution of Ar$^{4+}$ is negligible owing to its high IP (59.81 eV) and the non detection of {\heii} lines. We used the ICF shown in equation 36 by D-I14, which is valid when O$^{2+}$/(O$^{+}$+O$^{2+}$)$>$ 0.5. The uncertainties associated with this ICF are higher than for other elements (D-I14), but this ionization correction scheme is the best one available for this specie and our result is consistent with values from literature (see Table~\ref{abu_tot}).

We assumed Mg/H=Mg$^{2+}$/H$^+$ given the wide ionization potential interval occupied by Mg$^{2+}$ \citep{barlowetal03, wangliu07}. We obtain 12+log(Mg/H)=7.56$\pm$0.10 (see Table~\ref{abu_tot}). This value is compatible with the solar photospheric value of \citet{asplundetal09} (12+log(Mg/H)=7.60) and is identical to the average value of the Mg/H ratio in Galactic disk PNe \citep{wangliu07}. 

We used two different ICFs to calculate the total abundance of Fe, Equations~2 and 3 of \citet{rodriguezrubin05}. These two ICFs provide a range of Fe abundances. Additionally, we estimate the total Fe abundance from the sum of Fe$^+$ and Fe$^{2+}$ as 5.40$\pm$0.10. This value represents a lower limit of the total abundance of this element, since we ignore the contribution of \ffeiv, whose optical lines are intrinsically weak and were not detected.  Furthermore, the contribution of Fe$^+$ is uncertain, since its lines can arise both from the ionized portion of the nebula and from the neutral and partially molecular region. However, this sum agrees within the uncertainties with the lower limit of Fe abundance obtained from equation 3 of \citet{rodriguezrubin05}.
 
D-I14 do not provide corrections for unseen ions of P. In order to compute the total abundance of P, we took advantage of the similarity between the IP of P$^{+}$ and S$^{+}$ using the ionization correction scheme P/P$^{+}$=S/S$^{+}$. A detailed set of photoionization models is needed to find a more accurate ICF for this element.

\subsection{Neutron-capture element abundances}\label{sec:n_capt_abu}

The first ICF for Kr were provided by \citet{sterlingetal07}, through detailed photoionization models. Since then, atomic data for {\emph n}-capture ion transitions have considerably improved. New atomic data were incorporated in the new ICF computations for Kr and Se by \citet{sterlingetal15}. They performed a large grid of photoionization models in order to cover a considerable range of physical conditions. They derived 6 ICFs for Kr and 3 for Se, depending on which ions are detected. The quality of their Kr ICFs is confirmed by optical/IR data for 16 PNe \citep{sterlingetal15, garciarojasetal15}.  Thanks to the high ionization degree of the PN NGC\,3918, \citet{garciarojasetal15} detected several ionization stages of Kr in the optical spectrum. Therefore, they could use the complete set of ICFs provided by \citet{sterlingetal15} for Kr, finding very good agreement between all the results (see first row of their Table 17). Our results for NGC\,5315 are shown in Table~\ref{s_abu}. We choose the Kr abundance obtained with equation 4 of \citet{sterlingetal15}, because it incorporates both Kr$^{2+}$ and Kr$^{3+}$. Equation 1 provides a value that is very similar to the one selected. Equation 3 produces larger Kr abundances than the best empirical estimate when Ar$^{3+}$/Ar $<$ 0.20 \citep{sterlingetal15}, which is the case of NGC\,5315. The value provided by equation 2 is consistent within the uncertainties with those derive from equation 4, but it has large error bars, owing to the uncertainty in the Ar abundance.

Selenium abundances were computed using all the ICFs proposed by \citet{sterlingetal15}. The ICF of equation 7 is based only on Se$^{2+}$ ionic abundance, which is uncertain due to the weakness of the {\fseiii} 1.0992 $\mu$m line; Equation 8 provides an ICF that only considers Se$^{3+}$ and gives a value which is much lower than those provided by the other equations. \citet{sterlingetal15} argue that Equation 8 is the most uncertain of the Se ICFs, due to the lack of a strong correlation with fractional abundances of light element ions. 
We adopt the Se abundance derived with Equation 9 of \citet{sterlingetal15}, which is the most accurate since it takes into account both Se$^{2+}$ and Se$^{3+}$. 
Furthermore, the unblended line of {\fseiii} at 1.0992 $\mu$m, first identified by \citet{sterlingetal17}, allows for improved accuracy of Se abundances in PNe.  The Se abundances derived with the various ICFs of \citet{sterlingetal15} are shown in Table~\ref{s_abu}.

At present, there are no reliable ICFs for Xe, Br and Rb available in the literature. We calculated the total abundance of Xe using Equation 3 of \citet{sterlingetal15} as the ICF, given the similarity between the IPs of Xe$^{3+}$ and Kr$^{3+}$.  Our analysis yields a higher enrichment of Xe than for Kr (0.86~dex relative to Ar, compared to 0.02~dex for Kr), but we emphasize the uncertainty of the Xe abundance.  First, Equation~3 of \citet{sterlingetal15} gives systematically high Kr abundances in low- and moderate-excitation PNe (including NGC\,5315) compared to other ICFs.  We believe that this excitation effect plays a significant role in the anomalously large Xe abundance we derive.  Secondly, the {\fxeiv} $\lambda$7535.40 line is marginally detected, and {\fxeiv} $\lambda$5709.21 not at all, resulting in a quite uncertain Xe$^{3+}$/H$^+$ ionic abundance.

As we discussed in Section~\ref{sec:id}, we possibly detect {\fbriii} $\lambda$6556.56. Since the IP of Br$^{2+}$ is similar to the IP of Se$^{2+}$, we calculated the total abundance of Br by using equation 7 of \citet{sterlingetal15}. The Br abundance is shown in Table~\ref{abu_tot}. The flux of the line detected at 6555.56 $\AA$ gives a Br$^{2+}$ abundance much higher than expected, given the nearly solar abundances of the adjacent elements Se and Kr.
We inspected the spectra to check possible contamination or alternative identifications for the feature at 6555.56 $\AA$, but we did not find any transition with other detections belonging to the same multiplet in our spectra. We also computed an upper limit to the Br$^{2+}$/H$^+$ abundance from {\fbriii} $\lambda\sim$6130.40, which we did not detect.  This upper limit gives a bromine abundance $<$2.43, more than an order of magnitude lower than that derived from $\lambda$6556.56.  This result suggests that the {\fbriii} $\lambda$6556.56 is contaminated and/or incorrectly identified in NGC\,5315, and we conclude that our derived Br abundance is not reliable.

Finally, from an upper limit flux estimation of the {\frbiv} $\lambda$5759.55 line, we computed an upper limit (3$\sigma$) abundance for Rb$^{3+}$. Considering the similarity with the IP range of O$^{2+}$, we used the ionization correction scheme Rb/Rb$^{3+}$ = O/O$^{2+}$ \citep{sterlingetal16}. 

\setcounter{table}{12}
\begin{table}
\begin{minipage}{120 mm}
 \caption{Kr and Se total abundances (12 + log(X/H)).}
 \label{s_abu}
 \begin{tabular}{ccc}
  \hline
  ICF$^{a}$  &Kr & Se  \\
  \hline
  eq. 1  & 3.61$\pm$0.07 & --\\
  eq. 2 &   3.84$\pm$0.24 & -- \\                      
  eq. 3 &4.17$\pm$0.09 & --\\
   eq. 4 & 3.60$\pm$0.07 & -- \\
   eq. 7 & -- & 3.89$\pm$0.28 \\
     eq. 8 & -- & 3.13$\pm$0.11 \\
       eq. 9 & -- & 3.60$\pm$0.17 \\
         \hline
  \multicolumn{3}{l}{$^a$ ICFs provided by \citet{sterlingetal15}}\\
  \end{tabular}
  \end{minipage}
\end{table}

\begin{figure}
	\includegraphics[width=\columnwidth]{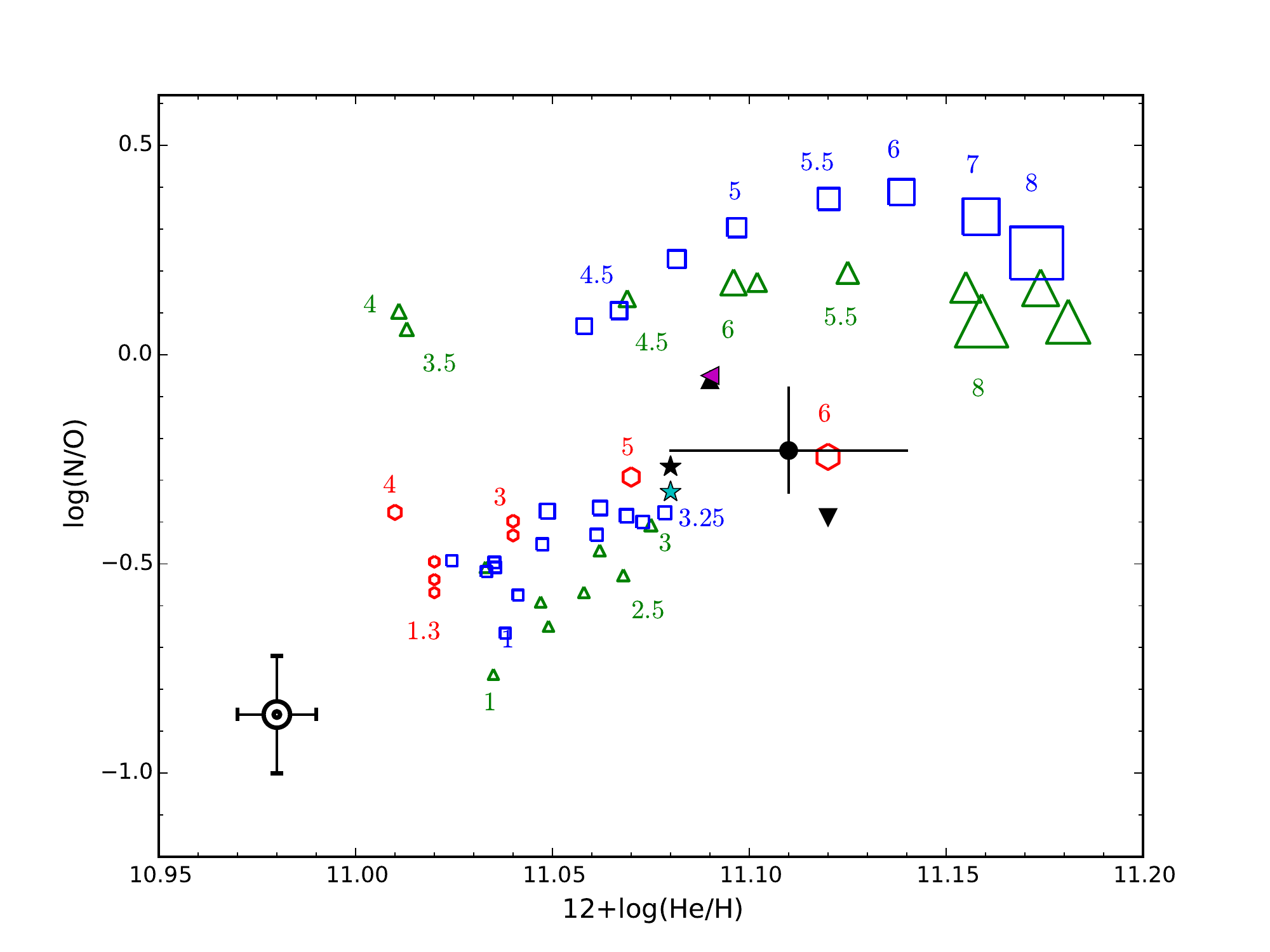}
	\includegraphics[width=\columnwidth]{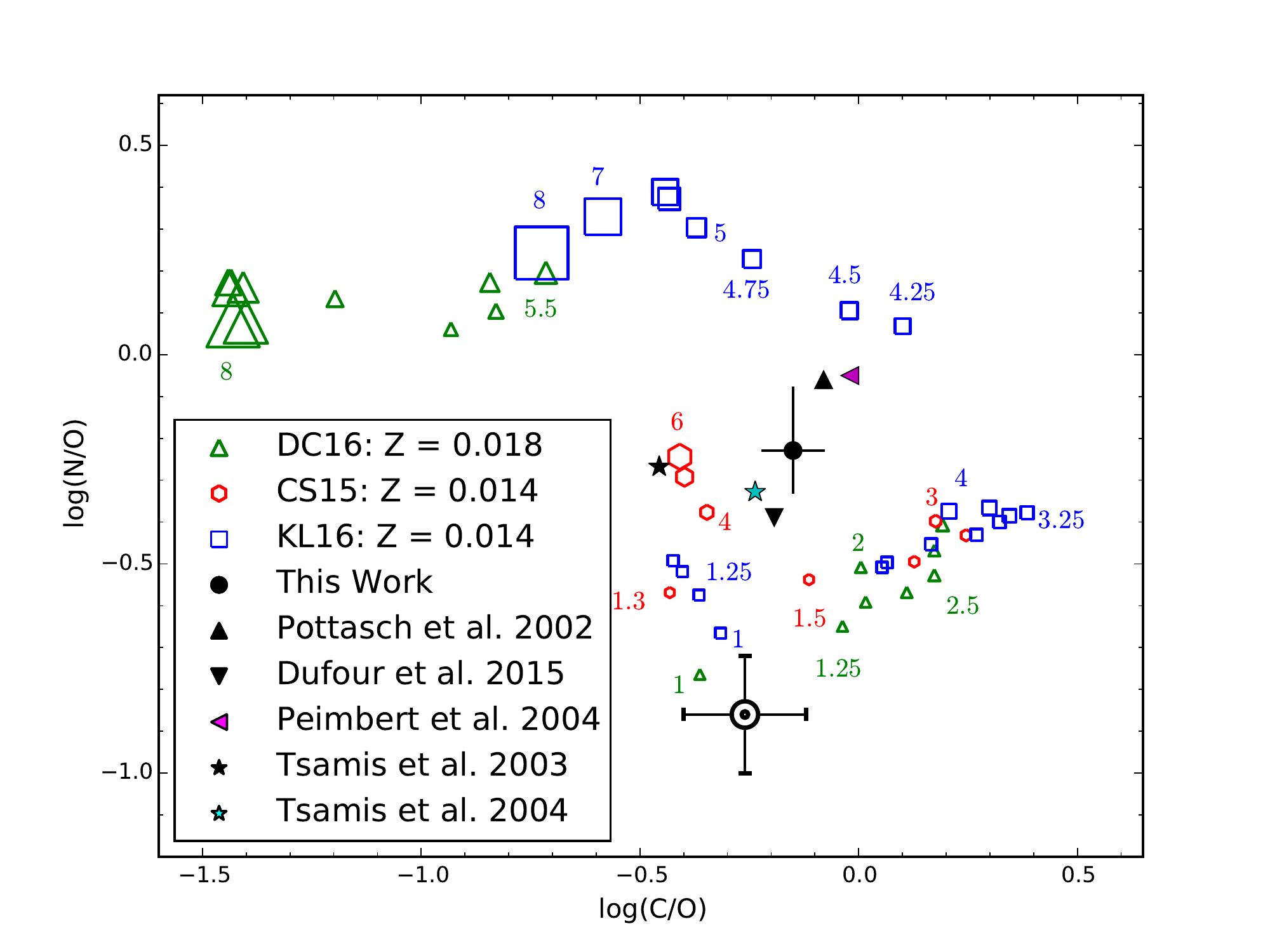}
    \caption{Upper panel: N/O vs. He/H ratios predicted by nucleosynthesis models. Lower panel: N/O  vs. C/O ratios. In both plots, blue squares are the values predicted by \citet{karakaslugaro16} nucleosynthesis models, green triangles from \citet{dicriscienzoetal16} nucleosynthesis models and red hexagons from \citet{cristalloetal15} nucleosynthesis models. Sizes of the symbols are scaled to the progenitor masses. Some labels have been included to help the reader to interpret the plots. The black dots are the observational results from our analysis. For comparison, we include abundance ratios from the literature.}
    \label{models}
\end{figure}


\setcounter{table}{13}
\begin{table}
        \caption{N/O and C/O ratios in NGC\,5315.}
        \label{n_o}
        \begin{tabular}{ccc} 
                \hline
                N/O& C/O & Reference\\
                \hline
                0.59$^{+0.24}_{-0.15}$ & 0.76$\pm$0.13 & This paper\\
                0.41 & 0.64 & \citet{dufouretal15}\\
                0.89 & -- & \citet{milingoetal10}\\
                0.93 & 0.95 & \citet{peimbertetal04}\\
                0.47 & 0.58 & \citet{tsamisetal04}\\
                0.54 & 0.35 & \citet{tsamisetal03}\\
                0.88 & 0.85 & \citet{pottaschetal02}\\
                0.48 & -- & \citet{liuetal01}\\
                0.59 & -- & \citet{samlandetal92}  \\
                0.36 & -- & \citet{defreitaspachecoetal91}\\
                \hline
        \end{tabular}
\end{table}

\setcounter{table}{14}
\begin{table*}
\begin{minipage}{120 mm}
	\caption{Neutron-Capture Element Abundances.}
	\label{s_enrich}
	\begin{tabular}{lcccc} 
		\hline
		&NGC\,5315 &NGC\,3918$^{a}$ &$<$Type I PNe$>$$^{b}$&$<$non-Type I PNe$>$$^{b}$\\										 
		\hline
		 $\lbrack$Se/(O,Ar)$\rbrack$& -0.07$\pm$0.29 & 0.19 &-0.01  & 0.20 \\
		 $\lbrack$Kr/(O,Ar)$\rbrack$& 0.02$\pm$0.26 &0.68& 0.13&0.91\\
		 $\lbrack$Br/(O,Ar)$\rbrack$& 0.66: & -- & -- & -- \\
		  $\lbrack$Rb/(O,Ar)$\rbrack$& $<$0.02 & $>$0.00 & -- & -- \\
		  $\lbrack$Xe/(O,Ar)$\rbrack$& 0.86: & $>$0.26 & -- & -- \\
                \hline
                                \multicolumn{5}{l}{$^a$ \citet{garciarojasetal15}}\\
               \multicolumn{5}{l}{$^b$ Mean values from the sample of \citet{sterlingetal15}}\\
                  	\end{tabular}
	\end{minipage}
\end{table*}

\begin{figure}
	\includegraphics[width=\columnwidth]{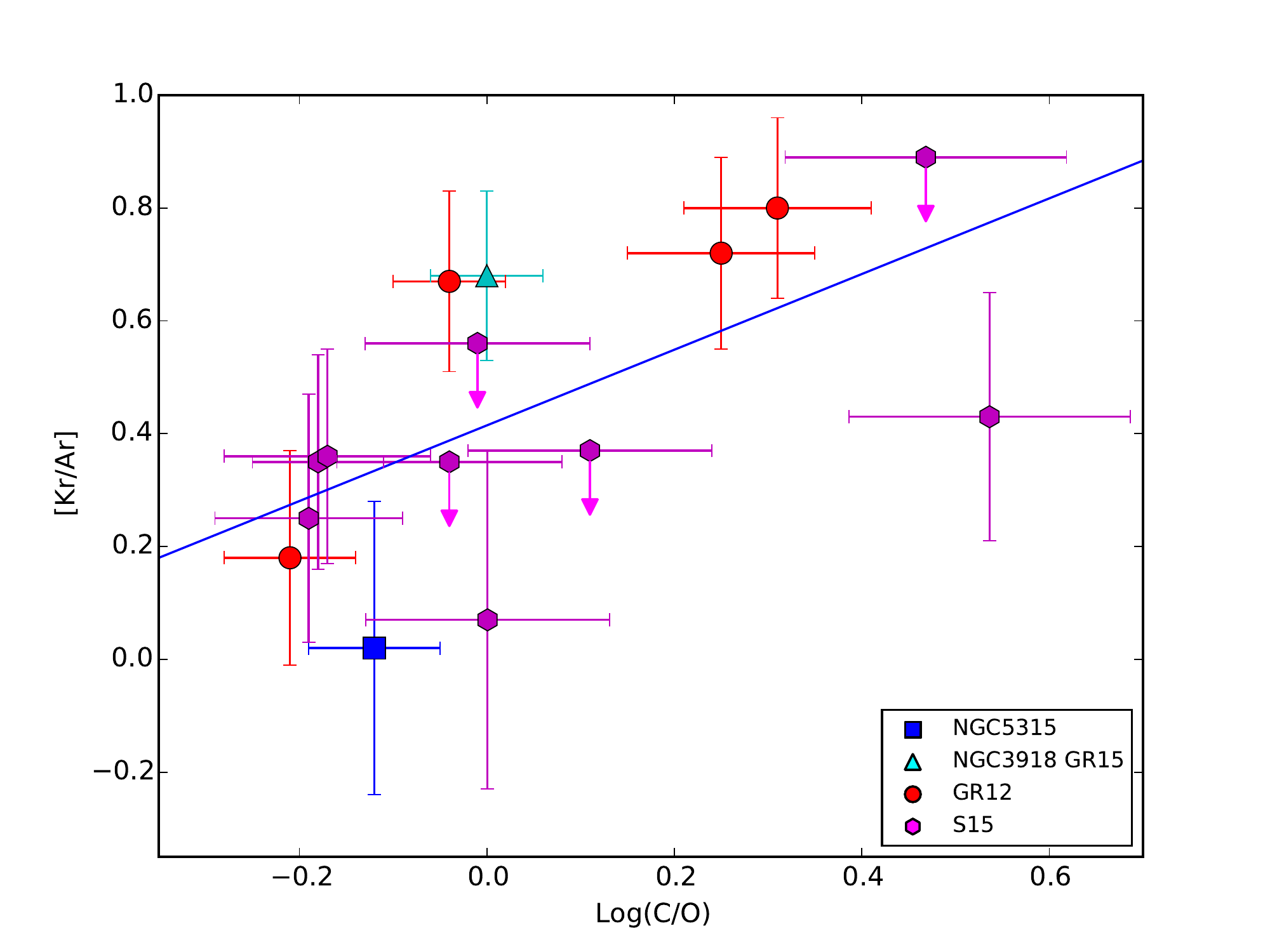}
    \caption{Correlation between Kr and C enrichment, using solar abundances by \citet{asplundetal09}. In this plot Kr enrichment is measured using Ar for all the PNe. The linear fit to the data is shown as a blue line. We have not considered upper limits to compute the fit. C/O ratios are computed from ORLs. Labels are the following: GR15: \citet{garciarojasetal15}; GR12: \citet{garciarojasetal12}; S15: \citet{sterlingetal15}.}
    \label{kr_c}
\end{figure}

\section{Discussion}\label{sec:discuss}

\subsection{CNO Abundances}\label{sec:n_o}

The N/O and the C/O ratios are crucial for investigating the nature of PN progenitor stars. In particular, the C/O ratio is an indicator of the chemistry of the nebula \citep[][and references therein]{sterlingdinerstein08} and it reveals crucial information about TDU events during the thermally-pulsing AGB.  The N/O ratio is a sensitive probe of HBB at the base of the convective envelope, which only occurs in stars with masses $\gtrsim 4$M$_{\odot}$.  In this section, we compare our determinations with those from the literature (see Table~\ref{n_o}) both to assess the accuracy of our abundances as well as to utilize information from other spectral regions.

Prior to our study, the deepest optical spectrum of NGC\,5315 was that analyzed by \citet{peimbertetal04}. They computed \hbox{N/O = 0.93} using a combination of ORLs and CELs, and \hbox{C/O = 0.93} from ORLs, assuming that temperature fluctuations are responsible for the ADFs. \citet{torrespeimbertpeimbert77} assumed the temperature fluctuations paradigm in their calculations and found \hbox{N/O = 0.95} from optical observations. Similarly high N/O ratios were found by \citet{pottaschetal02} and \citet{milingoetal10}.  In particular, \citet{pottaschetal02} used UV observations in which \niii$]$ and \niv$]$ were detected.  Along with optical data, these detections allowed them to compute the total N abundance by summing the N$^{+}$, N$^{2+}$ and N$^{3+}$ ionic abundances, eliminating the need for an ICF.

However, not all authors have found such large N/O ratios.  \citet{tsamisetal03} utilized \emph{IUE}, optical, and infrared data to compute \hbox{N/O = 0.54} from CELs, although their \hbox{C/O = 0.35} is lower than our derived value.  This group also determined abundances from ORLs \citep{tsamisetal04}, finding \hbox{N/O = 0.47} and \hbox{C/O = 0.58}.  \citet{dufouretal15} used long-slit $HST$-STIS co-spatial UV-optical spectra and computed \hbox{N/O = 0.41}, which is somewhat lower than our derived value and  \hbox{C/O = 0.64} which is consistent with our C/O ratio; the combination of optical and UV data allowed them to avoid using ICFs. \citet{liuetal01} found \hbox{N/O = 0.48} from a far-infrared \emph{ISO} spectrum.  Finally, \citet{defreitaspachecoetal91} and \citet{samlandetal92} derived \hbox{N/O = 0.36} and \hbox{N/O = 0.59}, respectively, from low-resolution optical spectra.

The discrepancy seen in the N/O ratios calculated in the literature can be ascribed to various factors, including the temperature sensitivity of UV \niii$]$ and \niv$]$ emission lines, and the large and uncertain ICFs required for optical spectra in which \fnii is the only N ion with CELs.  However, the average value calculated from the literature is \hbox{N/O = 0.63}, which is consistent within the uncertainties with our result \hbox{(N/O = 0.59$^{+0.24}_{-0.15}$)}. 

On the other hand C/O ratios from both ORLs and CELs seem to agree better, with the exception of the low value found by \citet{tsamisetal03} (Table~\ref{n_o}).  All extant C/O determinations are below unity, indicating that NGC\,5315 has an oxygen-rich chemistry.  Interestingly, this PN exhibits both O-rich crystalline silicate dust emission features and PAH features that are often associated with C-rich chemistries \citep{cohenetal05}.  This dual-dust chemistry has been found in other PNe with $[$WC$]$ central stars \citep[e.g.,][]{pereacalderonetal09} and may be associated with binary interactions \citep{demarcosoker02}.

Oxygen-rich PNe can be associated either with low mass stars (M $<$ 1.5 M$_{\odot}$), in which TDU events did not occur (preventing C enrichment on the stellar surface), or massive stars (M~$>$~3--4 M$_{\odot}$), where the temperature at the base of the convective envelope is high enough to active HBB.  In the latter case, C is converted to N by the CN cycle, leading to N and He enrichments.

Peimbert Type~I PNe exhibit N and/or He enrichments typical of HBB, and hence are believed to be descendants of more massive AGB stars.  This is supported by statistical evidence based on the scale heights, large central star masses and high central star temperatures of Type~I PNe \citep[e.~g.][]{peimbert90, corradischwarz95, gornyetal97, stanghellinietal02, penaetal13}.  Several different criteria have been proposed for Type~I classification in the literature.  The initial criterion defined by \citet{peimbert78} was overly restrictive because of the overestimation of He abundances and of electron temperatures which led to artificially large N/O ratios.  Therefore,  \citet{peimberttorrespeimbert83} refined these thresholds and defined Type I PNe as those having \hbox{He/H $>$ 0.125} or \hbox{N/O $>$ 0.5}. Later, \citet{macielquireza99} required that both of these thresholds be met for Type~I classification.  \citet{kingsburghbarlow94} proposed that Type~I PNe be classified on the basis of the N/O ratio alone, with the minimum N/O given by (C+N)/O in \hii regions of the host galaxy.  In the solar neighborhood, this limit is N/O $>$ 0.8.  Based on more recent \hii region and solar abundances, \citet{henryetal04} refined this criterion to N/O $\geq$ 0.65.  

Our He/H (0.129$\pm$0.009) and N/O (0.59$^{+0.24}_{-0.15}$) abundance determinations indicate that NGC\,5315 is a Type~I PN according to the \citet{macielquireza99} criterion, but not according to \citet{henryetal04}.  The uncertainties are sufficiently large that we cannot classify this PN either as a Type~I or non-Type~I with confidence.  Literature abundances are similarly inconclusive (Table~\ref{n_o}).  Nevertheless, it appears safe to say that NGC~5315 experienced some N and He enrichments, but HBB apparently did not operate efficiently in its progenitor.  This suggests that the progenitor of NGC~5315 may have had an initial mass M $\approx$ 3--5 M$_{\odot}$.  However, in this mass range the cessation of HBB can occur before the end of the thermally-pulsing AGB phase, and subsequent TDU episodes can lead to carbon enrichment \citep[e.g.,][]{frostetal98, venturaetal15}. 

To shed more light on the nature of the progenitor star, we compare our abundances with the nucleosynthetic models of \citet[][hereafter KL16]{karakaslugaro16}, \citet[][hereafter DC16]{dicriscienzoetal16} and \citet[][hereafter CS15]{cristalloetal15}. In Figure~\ref{models}, we show the predicted N/O vs. He/H ratios and N/O vs. C/O ratios for different progenitor masses with solar or near-solar metallicity, compared with our observed values. In the N/O vs He/H plot the mass of the progenitor star of NGC\,5315 is consistent, within the uncertainties, with predictions for 3--3.5 M$_{\odot}$ stars according to the predictions of KL16 and DC16. The CS15 models do not predict HBB at these masses at solar metallicities, and the predicted mass in this case is higher (5.5--6 M$_{\odot}$).  Literature abundances generally agree with this assessment, although the \citet{pottaschetal02} and \citet{peimbertetal04} data suggest a higher initial mass of 4.5--6.0 M$_{\odot}$ on account of their larger N/O ratios.  On the other hand, our N/O and C/O abundances (lower panel of Figure~\ref{models}) are not consistent with model predictions, except possibly for 1--1.5~M$_{\odot}$ values of KL16.  Literature abundances show a dichotomy, with those having larger N/O ratios closer to the 4--5~M$_{\odot}$ predictions of KL16, and those with smaller N/O agreeing more with 4--6~M$_{\odot}$ predictions from CS15 or KL16's 1--1.5~M$_{\odot}$ values.

Therefore light element abundances do not significantly constrain the progenitor mass of NGC\,5315, other than to suggest that an initial mass of 2--3~M$_{\odot}$ is unlikely.  It should be noted that while He and N are moderately enriched, as may be expected for a more massive progenitor, extra mixing processes in less massive stars during the RGB and AGB phases can also produce such enrichments \citep{nollettetal03}.

\subsection{Neutron-Capture Element Abundances}

The pattern of \emph{s}-process enrichments of \emph{n}-capture elements, or lack thereof, can also provide information about the nature of PN progenitor stars.  To assess whether the \emph{s}-process and TDU occurred during the AGB, it is necessary to assume an initial abundance pattern and to adopt a metallicity reference element.  For Galactic disk PNe such as NGC~5315, scaled solar abundances are usually a reasonable approximation \citep{henryetal04}.  Historically, O was used as reference element in non-Type I PNe, while in Type I PNe Ar turns out to be the best choice if O is affected by nucleosynthesis in the progenitor star of these objects \citep{sterlingdinerstein08, karakasetal09}. Moreover, \citet{delgadoingladaetal15} have found evidence for O enrichment in Galactic PNe with carbon-rich dust, suggesting that, in general, Ar or Cl are better reference elements. A positive value of [X/(O, Ar)]\footnotetext{The notation [X/Y] represents the logarithmic difference between nebular and solar ratio abundances, i.~e. [X/Y]=log(X/Y)-log(X/Y)$_{\odot}$.}, where X is a \emph{n}-capture element, indicates that the progenitor star experienced {\emph s}-process enrichment. We use \citet{asplundetal09} as the reference for the Solar abundances. 

In Table~\ref{s_enrich} we show the average value of Se and Kr enrichments found by  \citet{sterlingetal15} in their sample of Type I and non-Type I PN, compared with NGC\,5315 and \citet{garciarojasetal15}'s results for NGC\,3918. Given the difficulty of classifying NGC\,5315 as a Type~I or non-Type~I PN, we decided to calculate enrichments using Ar as reference element, and we chose the same reference for the sample of \citet{garciarojasetal12, garciarojasetal15}. We find that Kr and Se are not enriched, in contrast to the C-rich NGC~3918 \citet{garciarojasetal15}.

Rb is another key element for diagnosing \emph{s}-process nucleosynthesis, as it is an indicator of the main neutron source during the AGB phase. A large enrichment of Rb relative to other \emph{n}-capture elements indicates that $^{22}$Ne is the dominant reaction for the production of neutrons. In this case, the high neutron density activates different branching points in the \emph{s}-process path, increasing the production of Rb and Kr \citep{abiaetal01,vanraaietal12, karakasetal12}. According to models, stars with M $>$ 5 M$_{\odot}$ attain the temperatures required for the activation of the reaction $^{22}$Ne($\alpha$, \emph{n})$^{25}$Mg \citep{bussoetal99}, and thus Rb is an indicator of the mass of the progenitor star (see Sect.~\ref{sec:intro}).  We do not detect Rb in NGC\,5315, but compute a 3-$\sigma$ upper limit of [Rb/Ar]$<$0.02.  The lack of enrichment agrees with the Se and Kr abundances, and suggests that the progenitor of NGC\,5315 is less than 6~M$_{\odot}$ \citep{karakaslugaro16}. 

Our Br and Xe abundances (see Table~\ref{s_enrich}) must be taken with caution.  The identification of the $\lambda$6556 line as \fbriii is questionable, as discussed in \S \ref{sec:n_capt_abu}.  In the case of Xe, we marginally detected one line, {\fxeiv} $\lambda$7535.40.  For lack of Xe ICFs, we use Equation~3 of \citet{sterlingetal15} to compute the Xe abundance.  This ICF is known to produce systematically high ICFs in low- and moderate-excitation PNe, as was found specifically for NGC\,5315.  For this reason, we believe that the derived Xe abundance is overestimated by $\sim$0.60~dex, based on our results for Kr (Table~\ref{s_abu}).

The lack of \emph{s}-process enrichments indicate that the progenitor of NGC\,5315 either was a low-mass star ($M<1.5$~M$_{\odot}$) or an intermediate-mass star (3~M$_{\odot} < M < 6$~M$_{\odot}$).  

The lack of carbon enrichment is consistent with our derived {\emph n}-capture element abundances. Theoretical low-to-intermediate mass stellar evolution models predict a correlation between {\emph n}-capture element and C enrichment in PNe, as they are processed in the same stellar layers and dredged up together to the surface during TDU episodes \citep[][ KL14]{gallinoetal98, bussoetal01}. In Figure~\ref{kr_c} we plot the correlation between C/O and $[$Kr/Ar$]$ for PNe with the most accurate Kr abundances \citep[derived from multiple Kr ions, see][]{sterlingetal15}. We find evidence for a correlation, given by:
\begin{equation}
[{\rm Kr}/{\rm Ar}] = (0.472 \pm 0.093)  + (1.240 \pm 0.490) \log ({\rm C}/{\rm O}).
\end{equation}

New optical/NIR observations will be necessary to expand the sample and to strengthen this result, and search for correlations between C/O ratios and other {\emph n}-capture elements. 

\subsection{The Enigmatic Progenitor Star of NGC\,5315}

Armed with our abundance analysis, we now consider the various scenarios for the nature of the progenitor of NGC\,5315.  The result is rather unsatisfying, as we cannot discriminate between a low-mass and intermediate-mass origin for the progenitor.  The O-rich chemistry and lack of \emph{s}-process enrichment only seem to rule out the mass range 1.5~M$_{\odot} < M < 3.0$M$_{\odot}$, according to single star evolution.  The conclusions are even less restrictive if binary interactions led to the formation of NGC\,5315 -- although there is evidence that this indeed may be the case, as outlined below.

We find three viable evolutionary channels for the progenitor of NGC~5315.\\
\emph{i}) The intermediate-mass star scenario. NGC\,5315 exhibits moderate enrichments of He and N, as may be expected for a more massive progenitor star that underwent HBB.  However the N/O ratio appears to fall below the Type~I PN classification criteria, indicating that HBB did not modify its composition to a substantial extent.  The comparison between observed and predicted N/O and He abundances are consistent within the uncertainties with a progenitor mass of 3--6 M$_{\odot}$, depending on the AGB models used.  However, in the C/O vs.\ N/O plane, the measured abundances are consistent with an intermediate-mass progenitor only for the model predictions of \citet{cristalloetal15}.  Comparison with the \citet{karakaslugaro16} predictions instead indicate a low-mass progenitor ($M < 1.5$~M$_{\odot}$).  \citet{sterlingetal15} found a lack of \emph{s}-process enrichments of Se and Kr in Type~I PNe, perhaps due to the relatively small intershell mass and/or strong dilution of enriched material into the massive envelopes of the AGB progenitor stars.  The lack of \emph{s}-process enrichments in NGC\,5315 is consistent with this scenario.

\emph{ii}) The low-mass progenitor scenario.  The lack of \emph{s}-process enrichments and O-rich chemistry of the nebula can be equally well ascribed to a low-mass progenitor \citep[$M < 1.5$~M$_{\odot}$][and references therein]{karakaslugaro16} that did not experience the \emph{s}-process or TDU.  The He and N enrichments can be produced by extra mixing processes in low-mass stars during the RGB and the AGB phase \citep{nollettetal03}, although the nature of these phenomena are not well understood \citep[e. g.][]{karakaslugaro16}.  Additional support is provided by indirect estimates of the central star mass.  Using a luminosity based on the extinction distance, \citet{pottaschbernardsalas10} estimate M$\le$1.5 M$_{\odot}$ when comparing with post-AGB evolutionary tracks. From photoionization models, \citet{henryetal15} estimated the central star luminosity and used evolutionary tracks to determine a progenitor mass of 1.1 M$_{\odot}$.  Both of these mass estimates carry caveats.  The abundance analysis of \citet{pottaschbernardsalas10} indicated a much larger progenitor mass of M = 4.5 M$_{\odot}$.  The mass derived by \citet{henryetal15} is strongly model dependent, and their stellar luminosity disagrees both with statistical distance determinations and with the extinction distance of \citet{pottaschbernardsalas10}.  Further evidence against this scenario comes from $\alpha$-element abundances.  Because low-mass stars have longer lifetimes, the abundances of elements such as Cl and Ar should be lower than in Galactic {\hii} regions.   \citet{estebanetal15} determined 12+log ($<$Cl/H$>$)=5.09 and \citet{garciarojasesteban07} found 12+log($<$Ar/H$>$)=6.52 in \hii regions, which are lower than we find for NGC\,5315.  Furthermore, we compare our O, Cl and Ar abundances with the averaged values from a large sample of Type II PNe analyzed by \citet{kwitterhenry01}. They found lower abundances than our results for elements unaffected by stellar nucleosynthesis: 12+log ($<$O/H$>$)=8.74, 12+log ($<$Cl/H$>$)=5.28 and 12+log ($<$Ar/H$>$)=6.45, suggesting that the progenitor star of NGC\,5315 was formed more recently and hence is probably descent from a more massive progenitor star.

\emph{iii}) The binary progenitor scenario.  The peculiar multi-polar morphology of NGC\,5315 can be produced by binary interactions during the AGB \citep[][and references therein]{balickfrank02, demarco09, hillwigetal16}.  Mass transfer can reduce the envelope mass of the AGB star and prematurely truncate the AGB lifetime before significant C and \emph{s}-process enrichments can occur.  Depending on the orbit of the binary companion, the N and He enrichments could have been produced by HBB, or by extra mixing (or some other mixing process triggered by binary interactions) \citep{nollettetal03}.  This scenario is supported by the findings of \citet{manicketal15}, who conducted a detailed study of radial velocity (RV) variations of a sample of central stars of PNe. They found a strong RV variation for the central star of NGC\,5315, which may be explained with the presence of a close companion. \citet{manicketal15} emphasize that this result is preliminary because their observations only include four epochs, which is not sufficient to conclude that it has a periodic RV variation typical of binary systems.  A further complication is the Wolf-Rayet central star itself, whose stellar wind could produce RV variations.  A binary central star can also explain the dual-dust chemistry of NGC\,5315 and other PNe with $[$WC$]$ central stars \citep{demarcosoker02, cohenetal05, pereacalderonetal09}.  Finally, a binary origin could help to bridge the dissonance between progenitor mass estimates from abundances and from other methods \citep{pottaschbernardsalas10}.

We speculate that NGC\,5315 most likely had a binary progenitor star, although we cannot dismiss the first two scenarios.  The combination of RV variations, PAH and silicate dust emission features, and an overall lack of agreement between observed and predicted abundances support this scenario.  The moderate N and He enrichments and Rb upper limit suggest that NGC\,5315 could not have had a (single) progenitor mass above $\sim$5~M$_{\odot}$.  For stars in this mass range, HBB can cease before the thermally-pulsing AGB phase is over, allowing for TDU to produce C and \emph{s}-process enrichments \citep{frostetal98, venturaetal15}.  The $\alpha$-element abundances of NGC~5315 suggest that a low-mass progenitor ($M < 1.5$M$_{\odot}$, scenario~(\emph{ii})) is unlikely.

However, more data are needed to reveal the nature of NGC\,5315's progenitor star.  Specifically, observations are needed to test whether the RV variations found by \citet{manicketal15} are periodic.  If the central star indeed has a binary companion, its orbital characteristics can be used to constrain the likelihood of mass transfer during the AGB.  Furthermore, a detailed analysis and modeling of the $[$WC4$]$ central star \citep[e.g.][]{koesterkehamann97} can constrain the stellar luminosity and hence mass. \citet{marcolinoetal07} and \citet{todtetal15}used detailed stellar atmosphere codes which are appropriate when non-LTE conditions and stellar  winds are present. However, as the distance estimates to Galactic PNe are unreliable, they assumed a `standard'' luminosity for their models and scale their results to other luminosities through a transformed radius \citep[see][]{todtetal15}. The upcoming GAIA data release of precise parallaxes will turn around this behaviour. This information can be used to more accurately estimate the initial mass than has been possible.

These different scenarios highlight other open questions. The origin of non-spherical PNe has always been a matter of debate. There is still no quantitative theory to explain how a massive single star can form a highly non-spherical PN. The multipolar morphology can be explained with central binary systems, but only in a small fraction of them the orbital period is short enough to allow the companion to interact with the AGB star and modify the morphology of the PN \citep{demarcoizzard16}. Therefore, the small fraction of single massive stars and compact binary systems cannot explain the high concentration of non-spherical PNe ($\approx$ 80 $\%$; \citet{parkeretal06}). Another unresolved problem concerns the large percentage of Type I PNe ($\approx$ 20 $\%$; \citet{kingsburghbarlow94}). The IMF suggests that stars with M $>$ 3 M$_{\odot}$ are only a few percent of the total population of our Galaxy. Therefore, the question is: how many Type I PNe come from high mass progenitor stars? Rb abundance determinations in Type I PNe are urgently needed to verify which Type I PNe are descendants of the Rb-rich AGB stars observed by \citet{garciahernandezetal06, garciahernandezetal09}. Finally, a crucial point for future works is to investigate whether high N and He abundances can be understood in the framework of binary interactions and/or extra mixing in low (or intermediate) mass stars.

\section{Summary and conclusions}\label{sec:conclusions}

We analyzed the high-resolution (R$\sim$40000) optical spectrum obtained with UVES at the Very Large Telescope, and the medium-resolution (R$\sim$4800) near-infrared spectrum with FIRE at Magellan Baade telescope of NGC5315. We detect, identify and measure the intensities of about 700 lines in both spectra. Physical conditions were computed using intensity ratios of common ions. Ionic abundances were computed for ions which present available atomic data. Elemental abundances of common elements (O, Ar, Cl, N, S...) were performed using the more recent ICFs available when needed. The total abundances of \emph{n}-capture elements were calculated for Kr, Se, Xe, Br and Rb, using sophisticated ICFs for Kr and Se and the best approximate correction scheme for Xe, Br and Rb. Finally, the enrichment of these elements was measured, using Ar as reference element, against solar abundances.  

The aim of this work was to study the \emph{s}-process in the progenitor of the PN NGC\,5315. 
The relatively high concentration of N and He computed, may be associated to the occurrence of HBB and second dredge up, which are activated in AGB stars with M$>$3--4 M$_{\odot}$. Nevertheless, extra mixing events, thermohaline mixing \citep{eggletonetal06} and magnetic buoyancy \citep{trippellaetal14} in less massive stars during first and third dredge up may also enhance the final amount of N and He. The causes of these processes are still poorly known. 
Furthermore, the multipolar morphology of NGC\,5315 may be explained with the presence of a central binary system. 
Taking in account these considerations, we suggest that the most reasonable interpretation to the observed chemical abundances and for the absence of \emph{s}-process enrichment in NGC\,5315 is that a central binary system can be the cause of a poor or absent \emph{s}-process enrichment, due to a strong interaction with a close companion that can reduce the AGB lifetime through mass transfer. However, we cannot rule out other two alternative interpretations: i) that this PN descend from a progenitor star with mass between 4--6 M$_{\odot}$, where the massive convective envelope dilutes the \emph{n}-capture enrichment during TDU events and the intershell region between He- and H-burning shells is smaller than in lower mass stars; ii) finally, a low mass progenitor star (M$<$1.5 M$_{\odot}$) that did not experience TDU, and the stellar surface is not enriched in \emph{s}-processed material.

Our FIRE spectrum allows us to test the complete set of Se ICFs developed by \citet{sterlingetal15}. We consider the ICF given by their equation 9, which considers two different ions, as representative of Se total abundance. The identification of the non-blended {\fseiii} $\lambda$1.0995 $\mu$m line provides a much more robust test for the Se ICFs, than the blended {\fseiii} 8854 \AA\  line.

Finally, another important theoretical prediction was tested: the correlation between Kr and C enrichment. \emph{n}-capture elements are formed in the same layers of C, both are dredged up together to the stellar surface during TDU episodes and then stellar winds and PN ejections expel them in the interstellar medium. Therefore, we expect a correlation which is shown in Figure~\ref{kr_c}.

\section*{Acknowledgments}

This work is based on observations collected at the European Southern Observatory, Chile, proposal number ESO 092.D-0265(A).
This work has been funded by the Spanish Ministry of Economy and Competitiveness (MINECO) under the grants AYA2011-22614 and AYA2015-65205-P. SM acknowledges support of the Instituto de Astrof\'{\i}sica de Canarias under a PhD fellowship. JG-R acknowledges support from Severo Ochoa Excellence Program (SEV-2015-0548) Advanced Postdoctoral Fellowship. NCS acknowledges partial support from NSF award AST-0901432. GD-I gratefully acknowledges support from: Conacyt grant no. CB-2014/241732 and PAPIIT (DGAPA-UNAM) grants no. 107215 and IA-101517. SM
acknowledges the University of West Georgia, where part of this work was done. We thank D. A. Garc\'{\i}a-Hern\'andez for fruitful discussions. AMD acknowledges support from the FONDECYT project 3140383.  We thank the anonymous referee for a detailed report that help to improve the scientific content of the paper.






\onecolumn
\setcounter{table}{1}
\begin{longtable}{ccccccccc} 
\caption{Observed and reddening corrected line ratios (F(H$\beta$) = 100) and line identifications in NGC\,5315.\label{lineid_uves}}\tabularnewline
\hline\hline 
$\lambda_0$ (\AA\ ) & Ion & Mult. & $\lambda_{\rm obs}$ & $V_{\rm rad}$ (km s$^{-1}$) & $F(\lambda)/F({\rm H}ref)$$^{\rm d}$ & $I(\lambda)/I({\rm H}\beta)$$^{\rm e}$ & Err ($\%$)$^{f}$ & notes \tabularnewline
\hline
\endfirsthead
\caption{continued.}\tabularnewline
\hline\hline 
$\lambda_0$ (\AA\ ) & Ion & Mult. & $\lambda_{\rm obs}$ & $V_{\rm rad}$ (km s$^{-1}$) & $F(\lambda)/F({\rm H}ref)$$^{\rm d}$ & $I(\lambda)/I({\rm H}\beta)$$^{\rm e}$ & Err ($\%$)$^{f}$ & notes \tabularnewline
\hline
\endhead
\hline
\endfoot
\endlastfoot
3109.18	&	{\fariii}	&	2F	&	3108.70	&	-46.28	&	0.132	&	0.276	&	14	&		\\
3187.84	&	{\hei}	&	3	&	3187.31	&	-49.84	&	1.240	&	2.453	&	10	&		\\
3218.19	&	{\neii}	&	4D$_0$-4F	&	3217.71	&	-44.71	&	0.0741	&	0.144	&	16	&		\\
3244.00	&	{\oii}	&	4P-D[2]$_{0}$	&	3243.40	&	-55.45	&	0.0615	&	0.117	&	18	&	?	\\
3258.27	&	{\hei}	&	1S-1P$_{0}$	&	3257.79	&	-44.16	&	0.0401	&	0.0757	&	22	&		\\
3296.77	&	{\hei}	&	9	&	3296.22	&	-50.01	&	0.0363	&	0.0671	&	23	&		\\
3323.74	&	{\neii}	&	7	&	3323.32	&	-37.88	&	0.0287	&	0.0524	&	27	&		\\
3328.72	&	{\nii}	&	3D-3P$_{0}$	&	3328.39	&	-29.72	&	0.0461	&	0.0839	&	20	&		\\
\end{longtable}

\clearpage

\twocolumn

\onecolumn
\setcounter{table}{2}
\begin{longtable}{ccccccccc} 
\caption{Observed and reddening corrected line ratios (F(H$\beta$) = 100) and line identifications in the FIRE spectrum of NGC\,5315.\label{lineid_fire}}\tabularnewline
\hline\hline 
$\lambda_0$ (\AA\ ) & Ion & Mult. & $\lambda_{\rm obs}$ & $V_{\rm rad}$ (km s$^{-1}$) & $F(\lambda)/F({\rm H}ref)$$^{\rm d}$ & $I(\lambda)/I({\rm H}\beta)$$^{\rm e}$ & Err ($\%$)$^{f}$& notes \tabularnewline
\hline
\endfirsthead
\caption{continued.}\tabularnewline
\hline\hline 
$\lambda_0$ (\AA\ ) & Ion & Mult. & $\lambda_{\rm obs}$ & $V_{\rm rad}$ (km s$^{-1}$) & $F(\lambda)/F({\rm H}ref)$$^{\rm d}$ & $I(\lambda)/I({\rm H}\beta)$$^{\rm e}$ & Err ($\%$)$^{f}$& notes \tabularnewline
\hline
\endhead
\hline
\endfoot
\endlastfoot
8316.55	&	{\hi}	&	P26	&	8315.52	&	-37.07	&	2.772	&	0.162	&	10	&		\\
8325.71	&	{\hi}	&	P25	&	8324.70	&	-36.32	&	3.815	&	0.223	&	9	&		\\
8336.07	&	{\hi}	&	P24	&	8335.18	&	-31.99	&	3.937	&	0.230	&	9	&		\\
8347.84	&	{\hi}	&	P23	&	8346.82	&	-36.95	&	5.362	&	0.313	&	8	&		\\
8361.30	&	{\hi}	&	P22	&	8361.31	&	0.27	&	7.923	&	0.463	&	8	&		\\
8364.04	&	{\hei}	&	68	&	*	&	*	&	*	&	*	&	*	&		\\
8376.78	&	{\hi}	&	P21	&	8376.25	&	-19.13	&	6.377	&	0.372	&	8	&		\\
8394.71	&	{\hi}	&	P20	&	8394.08	&	-22.55	&	5.749	&	0.335	&	8	&		\\
\end{longtable}

\clearpage

\twocolumn







\bsp	
\label{lastpage}
\end{document}